\documentclass[aip,jcp,amsmath,amssymb, reprint]{revtex4-1}

\usepackage{braket}
\usepackage{amsthm}
\usepackage{graphicx}
\theoremstyle{remark}
\newtheorem{remark}{Remark}

\DeclareMathOperator{\Tr}{Tr}

\begin{document}


\title{Calculation of coherences in F\"orster and modified Redfield theories of excitation energy transfer}



\author{Anton Trushechkin}
\email{trushechkin@mi-ras.ru}
 \affiliation{Steklov Mathematical Institute of Russian Academy of Sciences, Gubkina 8, 119991 Moscow, Russia}
 \affiliation{National Research Nuclear University MEPhI, Kashirskoe highway 31, 115409 Moscow, Russia}
 \affiliation{National  University of Science and Technology MISIS, Leninsky avenue 2, 119049 Moscow, Russia}


\date{\today}

\begin{abstract}
F\"orster and modified Redfield theories play one of the central roles in the description of excitation energy transfer in molecular systems. However, in the present state, these theories  describe only the dynamics of populations of local electronic excitations or delocalized exciton eigenstates, respectively, i.e., the diagonal elements of the density matrix in the corresponding representation. They do not give prescription for propagating the off-diagonal elements of the density matrix (coherences). This is commonly accepted as a limitation of these theories. Here we derive formulas for the dynamics of the coherences in the framework of F\"orster and modified Redfield theories and, thus, remove this limitation. These formulas provide excellent correspondence with numerically exact calculations according to the hierarchical equations of motion. Also we show that, even within the range of applicability of the standard Redfield theory, the formulas for coherences derived in the framework of the modified Redfield theory provide, in some cases, more precise results.
\end{abstract}


\maketitle 

\section{Introduction}

Mathematical description of excitation energy transfer (EET) in molecular systems\cite{MayKuhn,Valkunas,RengerMayKuhn} is an important branch of theory of open quantum systems and chemical physics. This field of research finds applications in quantum technologies and biological systems. An important example of the latter is light-harvesting pigment-protein complexes in photosynthetic systems of plants and some bacteria.\cite{QEffBio} These complexes are responsible for the absorption of light and the transfer of the excitation energy into photosynthetic reaction centers. Recently, experimental evidences of long-lived quantum coherences in EET  in light-harvesting complexes were obtained.\cite{Engel,Scholes}. This rises  questions about the origin of these long-lived coherences and their possible role in EET. Note that EET  in light-harvesting complexes is highly efficient: usually, more than 95\% of the absorbed quanta of energy  are transferred to  reaction centers.\cite{QEffBio}

From the viewpoint of theory of open quantum systems,\cite{MayKuhn,Valkunas,BP,AccLuVol,AccKozLec,Huelga} the electronic degrees of freedom of molecules constitute ``a system'', which is coupled to ``a bath'' consisting of the vibrational degrees of freedom and the environment of the molecules. Quantum master equations for either the whole reduced density matrix of the system or only its diagonal part in some basis, is a widely used tool of theory of open quantum system.

There are three main approaches (or theories) leading to Markovian quantum master equations. If the dipole couplings between different molecules can be treated as small parameters, then the corresponding perturbation theory leads to  F\"orster resonance energy transfer theory.\cite{Forster1,Forster2} On the other hand, treating the system-bath coupling as a small perturbation parameter leads to theory of weak-coupling limit, or Redfield theory (which is often referred to as standard Redfield theory).\cite{Redfield,Davies,AccLuVol,AccKozLec} Alternatively, treating only the off-diagonal part of the system-bath interaction Hamiltonian (in the exciton representation) leads to the modified Redfield theory.\cite{Zhang,YangFl}. 

All three approaches have shown to be valid for the description of EET  in molecular systems in certain regimes and provide useful insights into our understanding of these phenomena. In Ref.~\onlinecite{Seibt} it is stressed that, despite of theoretical development of more advanced models (for example, with distinguished vibrational modes\cite{Kolli,PlenioOrigin,novo2017}) and non-Markovian master equations (for example, the hierarchical equations of motions\cite{IFl} or the polaron-representation master equation\cite{Polaron0,Polaron}), these three theories will continue to play an important role for the description of EET for years to come.

Unfortunately, unlike the standard Redfield theory, F\"orster and the modified Redfield theories describe the dynamics of only the diagonal part of the reduced density matrix of the system (populations) in the local electronic basis and exciton eigenstate basis, respectively. They do not describe the dynamics for the off-diagonal elements of the reduced density matrix (coherences). This is commonly accepted as a  limitation of these theories.\cite{YangFl,IFlRedf,NovoGrond,Valkunas,Seibt} In Ref.~\onlinecite{YangFl} it is stressed that the population transfer  rates are not enough for  the description of optical experiments and the off-diagonal elements are needed. 

Here we derive formulas for the dynamics of coherences in the framework of  F\"orster and modified Redfield theories and, thus, remove this limitation. Our formulas provide  good correspondence with the numerically exact method of hierarchical equations of motion (HEOM).\cite{IFl} Also we show that modified Redfield theory is at least as good as standard Redfield theory in predicting the exciton coherences and local site populations and, in some cases (even within the range of validity of the standard Redfield equation), provides more accurate results.

Note that, in Ref.~\onlinecite{HwangFu}, a phenomenological equation for the dynamics of coherences within modified Redfield theory is proposed. According to this equation, the coherences exponentially decay to zero. In contrast, we derive formulas for coherences using a rigorous method. Namely, we use the Zwanzig projector (super)operator method.\cite{Zwanzig} Usually it is used for derivation of quantum master equations. But, in fact, it can be used for more general purposes as well. We show that the coherences do not decay to zero, but tend to some stationary non-zero values. This is confirmed by the HEOM method as well.

The paper is organized as follows. The theoretical background is reviewed in Sec.~\ref{SecRev}. It includes the Hamiltonian (Sec.~\ref{SecHam}), a general scheme of F\"orster and modified Redfield theories in terms of the Zwanzig projection operator formalism (Sec.~\ref{SecQME}) and the application of this general scheme to the particular cases of F\"orster (Sec.~\ref{SecForst}) and modified Redfield (Sec.~\ref{SecMRedf}) theories. 

The new formulas for coherences are derived in Sec.~\ref{SecCoh}. Namely, subsection~\ref{SecEq} is devoted to the case of the equilibrium initial state of the bath and no initial coherences. In subsection~\ref{SecNoneq},  the case of a non-equilibrium initial state of the bath is considered. In  subsection~\ref{SecIni}, these results are generalized to the case of an arbitrary initial density matrix, i.e., with initial coherences. Every subsection in Sec.~\ref{SecCoh} includes  formulas in the framework of the general scheme and their application to  F\"orster and modified Redfield theories. In Sec.~\ref{SecCalc}, we provide examples of calculations of  coherences according to the derived formulas and compare them with the numerically exact  HEOM method. 

After that, in Sec.~\ref{SecDiscus}, we discuss several issues. The developed approach allows to calculate the evolution of the whole electronic density matrix (not just its diagonal elements) for an arbitrary initial electronic density matrix (i.e., also not necessarily diagonal). In subsection~\ref{SecMap} we discuss the properties of the corresponding dynamical map $\Lambda_t$, which maps the initial electronic density matrix $\sigma(0)$ to the evolved one $\sigma(t)$. In particular, we argue that the dynamics of the whole density matrix can be treated as non-Markovian, despite the fact that the dynamics of the populations alone is Markovian. In subsection~\ref{SecHop} we discuss in which sense the F\"orster mechanism of EET can be said to be ``incoherent hopping''. Finally, in subsection~\ref{SecVal}, we give analytical rough estimates of the magnitude of electronic coherences in the F\"orster approach and discuss the range of validity of this approach. In particular, we show that the F\"orster approach may be adequate even if the intersite Coulombic coupling is much larger than the reorganization energy.

\section{Theoretical background}\label{SecRev}
\subsection{Hamiltonian}\label{SecHam}

The Hamiltonian describing EET processes in molecular aggregates is as follows (the so called Frenkel exciton Hamiltonian):\cite{YangFl,QEffBio,Seibt}

\begin{subequations}
\begin{eqnarray}
&&H=H^{\rm el}+H^{\rm Coul}+H^{\rm ph}+H^{\text{el-ph}}+
H^{\rm reorg},\\
&&H^{\rm el}=\sum_{n=1}^N\ket n\varepsilon^0_n\bra n,\\
&&H^{\rm Coul}=\sum_{n=1}^N\sum_{m>n}^N(J_{nm}\ket n\bra m+\text{h.c.}),\\
&&H^{\rm ph}=\sum_{n=1}^NH^{\rm ph}_n,\\
&&H^{\rm ph}_n=\sum_i
\left(
\frac{p_{ni}^2}{2M_{ni}}+
\frac12M_{ni}\omega_{ni}^2q_{ni}^2
\right),\\
&&H^{\text{el-ph}}=\sum_{n=1}^N\ket n u_n\bra n,
\:\,
u_n=\sum_iM_{ni}\omega_{ni}^2d_{ni}q_{ni},\\
&&H^{\rm reorg}=\sum_{n=1}^N\ket n \lambda_n\bra n,
\:\,\lambda_n=\frac12\sum_iM_{ni}\omega_{ni}^2d_{ni}^2,
\end{eqnarray}
\end{subequations}
where ``h.c.'' stands for Hermitian conjugate. Here the terms $H^{\rm el}$ and $H^{\rm Coul}$ represent the electronic (system) part: $N$ is the number of monomers (e.g., individual molecules) in the aggregate, $\ket n$ represents the excited electronic state of the $n$th monomer (with all other monomers being in the ground state),  $\varepsilon^0_n$ is the electronic excitation energy of the $n$th monomer, $J_{nm}$ is the dipole Coulombic coupling constant between the $n$th and $m$th monomers. These coupling constants are responsible for EET between the monomers. Each monomer $n$ is coupled to its own phononic bath consisting of harmonic oscillators, with $q_{ni}$ and $p_{ni}$ being the position and momentum operators of the $i$th phonon mode of the corresponding bath. The parameters $M_{ni}$ and $\omega_{ni}$ are the mass and frequency of the corresponding mode, and $d_{ni}$ is the displacement of the equilibrium configuration of the mode between the ground and excited electronic states of the monomer. These displacements $d_{ni}$ play the role of coupling constants between the system (electronic degrees of freedom) and the bath (phononic degrees of freedom). The system-bath interaction manifests itself in the interaction  Hamiltonian $H^{\text{el-ph}}$ and also in renormalization of the electronic excitation energies described by the term $H^{\rm reorg}$, with $\lambda_n$ being the reorganization energy of the $n$th monomer.

The dynamics of the density matrix $\rho(t)$ of both electronic and phononic degrees of freedom is given by the von Neumann equation
\begin{equation}\label{EqNeumann}
\dot\rho(t)=-i[H,\rho(t)].
\end{equation}

\subsection{General scheme of F\"orster and modified Redfield theories}\label{SecQME}

Since the bath part includes an infinite number of degrees of freedom, even numerical solution of equation (\ref{EqNeumann}) is challenging. Perturbation theory and derivation of  quantum master equations for a finite number of degrees of freedom is one of the main tools of theory of open quantum systems. The Zwanzig projection superoperator method is widely used for derivation of quantum master equations. It is based on distinguishing between ``slow'' and ``fast'' degrees of freedom. Here we describe a particular application of this method. We follow Refs.~\onlinecite{YangFl} and~\onlinecite{Seibt}. Let $\{\ket\alpha\}$ be an orthonormal basis of the system such that the off-diagonal part of the Hamiltonian $\braket{\alpha|H|\beta}$, $\alpha\neq\beta$, can be treated as a small perturbation. Then we can divide the Hamiltonian into two parts (reference and perturbation Hamiltonians):
\begin{equation}\label{EqPert}
H=H_0+H',
\end{equation}
where
\begin{subequations}\label{EqH0Hpr}
\begin{eqnarray}
&&H_0=\sum_\alpha\ket\alpha H_\alpha\bra\alpha,\quad 
H_\alpha=\braket{\alpha|H|\alpha},\\
&&H'=\sum_{\alpha\neq\beta}\ket\alpha H'_{\alpha\beta}\bra\beta,
\quad 
H'_{\alpha\beta}=\braket{\alpha|H|\beta}.\label{EqHpr}
\end{eqnarray}
\end{subequations}

We assume the the evolution generated by the unperturbed part $H_0$ can be treated exactly. In the interaction representation with respect to $H_0$, we have
\begin{equation}\label{EqLV}
\dot\rho^{(I)}(t)=
-i[V(t),\rho^{(I)}(t)]\equiv
-i\mathcal L(t)\rho^{(I)}(t),
\end{equation}
where $\rho^{(I)}(t)=e^{iH_0t}\rho(t)e^{-iH_0t}$,
 $\mathcal L(t)=[H'(t),\cdot]$, 
$H'(t)=e^{iH_0t}H'e^{-iH_0t}.$
In the following, we will always work in the interaction representation and will omit  the superscript $(I)$. 

Let us introduce a projection (super)operator $\mathcal P$ acting on an arbitrary operator $A$ on the whole space of electronic and phononic degrees of freedom as follows:
\begin{equation}\label{EqP}
\mathcal PA=\sum_\alpha\Tr_B
(A_{\alpha\alpha})\ket\alpha\bra\alpha\rho_\alpha,
\quad A_{\alpha\alpha}=\braket{\alpha|A|\alpha}.
\end{equation}
Here $\Tr_B$ is the partial trace over the bath (phonons), and $\rho_\alpha=e^{-\beta H_\alpha}/\Tr e^{-\beta H_\alpha}$ is the equilibrium state of the bath corresponding to the system state $\ket\alpha$ (here $\beta$ is the inverse temperature, not to be confused with the index $\beta$ in Eq.~(\ref{EqH0Hpr})). The two additional properties for such projection operator are satisfied:
\begin{equation}\label{EqPLP}
[\mathcal P,\mathcal L_0]=0,\quad
\mathcal{PL}(t)\mathcal P=0,
\end{equation}
where $\mathcal L_0=[H_0,\cdot]$.

Let us also denote $\mathcal Q=1-\mathcal P$. The density operator $\rho(t)$, thus, can be divided into two parts:
\begin{equation}
\rho(t)=\mathcal P\rho(t)+\mathcal Q\rho(t).
\end{equation}
Inserting the resolution of identity $\mathcal P+\mathcal Q$ in front of $\rho(t)$ in the right-hand side of Eq.~(\ref{EqLV}) yields the system of equations
\begin{subequations}\label{EqPQrho}
\begin{eqnarray}
\mathcal P\dot\rho(t)&=&-i\mathcal P\mathcal L(t)\mathcal Q\rho(t),\label{EqPrho}\\
\mathcal Q\dot\rho(t)&=&
-i\mathcal L(t)\mathcal P\rho(t)
-i\mathcal Q\mathcal L(t)\mathcal Q\rho(t),\label{EqQrho}
\end{eqnarray}
\end{subequations}
where properties (\ref{EqPLP})  were used.

If we treat $\mathcal P\rho(t)$ as a known function, then a formal solution of  equation (\ref{EqQrho}) for $\mathcal Q\rho(t)$ is:
\begin{multline}\label{EqQ}
\mathcal Q\rho(t)=
{\rm T}_+\exp\left\lbrace
-i\int_0^t\mathcal Q\mathcal L(\tau)d\tau
\right\rbrace \mathcal Q\rho(0)\\-
i\int_0^t \,d\tau\,{\rm T}_+\exp
\left\lbrace
-i\int_\tau^t\mathcal Q\mathcal L(\tau')d\tau'
\right\rbrace\mathcal L(\tau) \mathcal P\rho(\tau),
\end{multline}
where 
\begin{multline}
{\rm T}_+\exp\left\lbrace
-i\int_0^t f(\tau)d\tau\right\rbrace\\=
1+\sum_{n=1}^\infty(-i)^n
\int_0^td\tau_1\int_0^{\tau_1}d\tau_2\ldots
\int_0^{\tau_{n-1}}d\tau_n\\
 f(\tau_1)f(\tau_2)\cdots f(\tau_n)
\end{multline}
is the chronological exponential. If we now substitute the formal solution (\ref{EqQ}) into Eq.~(\ref{EqPrho}), we will obtain the so called Nakajima--Zwanzig equation for $\mathcal P\rho(t)$. It is formally exact and equivalent to the von Neumann equation (\ref{EqLV}). But we are going to derive an approximate master equation for $\mathcal P\rho$ using the perturbation theory with respect to $H'$.

The operator $\mathcal P$ projects onto the subspace of slow degrees of freedom: as we see from Eq.~(\ref{EqPrho}), $\mathcal P\rho$ is changed slowly if $H'$ is small. It is assumed that the free (unperturbed) dynamics generated by $H_0$ moves $\mathcal Q\rho$ to zero: $e^{-iH_0t}\mathcal Q\rho e^{iH_0t}\to0$ as $t\to\infty$ in some sense. This condition (or its analogue) is required for the perturbation theory to be valid on arbitrary large times and, explicitly or implicitly, is used in derivations of master equations. But establishing a precise mathematical formulation of this condition and a rigorous proof of it is still desired. In our case, this condition means that the bath  relaxes to the state $\rho_\alpha$ provided that the system is in the state $\ket\alpha$, and the off-diagonal elements $\braket{\alpha|\rho|\beta}$, $\alpha\neq\beta$, disappear under the free dynamics. This is a fast relaxation process.

As we see from Eq.~(\ref{EqQrho}), the perturbation Hamiltonian $H'$ slowly moves $\mathcal Q\rho(t)$ away from zero. The smallness of the perturbation with respect to the rate of the bath relaxation assures the smallness of  $\mathcal Q\rho(t)$ for all times. This allows to use the perturbation theory with respect to $H'$ and leave only the lowest-order terms in Eq.~(\ref{EqQ}) for  arbitrarily large times.

Often it is said that $\mathcal P$ and $\mathcal Q$ project onto the subspaces of ``relevant'' and ``irrelevant'' degrees of freedom, but this is not always true: for example, here $\mathcal Q\rho$ includes the coherences $\braket{\alpha|\rho|\beta}$, $\alpha\neq\beta$, which are often of interest as well.  From the mathematical point of view it is better to say about ``slow'' and ``fast'' degrees of freedom. The separation between the time scales of the slow dynamics induced by $H'$ and the fast relaxation dynamics induced by $H_0$ assures the validity of perturbation theory. We discuss this in more detail in Sec.~\ref{SecVal}. 

Let, for simplicity, $\mathcal Q\rho(0)=0$. From Eq.~(\ref{EqQ}), in the first order with respect to $\mathcal L(t)$, we have
\begin{equation}\label{EqQ1}
\mathcal Q\rho(t)=
-i\int_0^t \mathcal L(\tau) \mathcal P\rho(\tau)\,d\tau.
\end{equation}
The substitution of this formula into Eq.~(\ref{EqPrho}) yields
\begin{equation}\label{EqNonMark}
\mathcal P\dot\rho(t)=
-
\int_0^t d\tau\,
\mathcal P\mathcal L(\tau)\mathcal L(0)\mathcal P\rho(t-\tau).
\end{equation}
Equation (\ref{EqNonMark}) is a non-Markovian quantum master equation for $\mathcal P\rho$. Here the non-Markovianity (see, e.g., Refs.~\onlinecite{BP} and~\onlinecite{Huelga} for more details) manifests itself as a dependence of the right-hand side not only on the current state $\mathcal P\rho(t)$ but also on the past states: $\mathcal P\rho(t-\tau)$.

Under some conditions, we can pass to a Markovian quantum master equation. The operator $\mathcal P$ includes the partial trace over the bath, which contains an infinite number of degrees of freedom. So, the integrand in Eq.~(\ref{EqNonMark}) reflects the fast bath relaxation processes and is expected to decay to zero much faster than $\mathcal P\rho$ evolves. In this case we can replace $\mathcal P\rho(t-\tau)$  by $\mathcal P\rho(t)$ in the right-hand side of  Eq.~(\ref{EqNonMark}) and extend the upper limit of integration to infinity there. This is the so called Born--Markov approximation,\cite{BP,Valkunas} which leads to the Markovian quantum master equation
\begin{equation}\label{EqMark}
\mathcal P\dot\rho(t)=\left\lbrace-
\int_0^\infty d\tau\,
\mathcal P\mathcal L(\tau)\mathcal L(0)\right\rbrace\mathcal P\rho(t).
\end{equation}

The substitution of  Eq.~(\ref{EqHpr}) and the expression
\begin{equation}
\mathcal P\rho(t)=\sum_\alpha p_\alpha(t)\ket\alpha\bra\alpha\rho_\alpha,
\end{equation}
where
\begin{equation}
 p_\alpha(t)=\Tr_B\braket{\alpha|\rho(t)|\alpha},
\end{equation}
 into Eq.~(\ref{EqMark}) yields the following master equations for $p_\alpha(t)$:

\begin{equation}\label{EqMaster}
\dot p_\alpha(t)=\sum_{\beta\neq\alpha}
[K_{\alpha\beta}p_\beta(t)-K_{\beta\alpha}p_\alpha(t)],
\end{equation}
where the transfer rates $K_{\alpha\beta}$ are given by
\begin{equation}\label{Eqk}
K_{\alpha\beta}=2\Re\int_0^\infty d\tau\,
\Tr[e^{iH_\beta\tau} H'_{\beta\alpha}e^{-iH_\alpha\tau}H'_{\alpha\beta}\rho_\beta].
\end{equation}

Note that the condition $\mathcal Q\rho(0)=0$ means that the initial state has the form
\begin{equation}\label{EqRhoEq}
\rho(0)=\sum_\alpha p_\alpha(0)\ket\alpha\bra\alpha\rho_\alpha,
\end{equation}
i.e., there are no initial coherences (off-diagonal parts of the density matrix) and the bath is in equilibrium (adjusted to the corresponding states of the system). A generalization of this scheme to a non-equilibrium case, where the initial state has the form
\begin{equation}\label{EqRhoNoneq}
\rho(0)=\sum_\alpha p_\alpha(0)\ket\alpha\bra\alpha\rho_g,
\end{equation}
where $\rho_g=e^{-\beta H^{\rm ph}}/\Tr e^{-\beta H^{\rm ph}}$ is given in Ref.~\onlinecite{Seibt}. However, if the bath relaxation is much faster than the population transfer, then the non-equilibrium corrections are negligible and we can still use Eq.~(\ref{Eqk}). For simplicity, we will assume this case and use Eq.~(\ref{Eqk}) for population transfer rates.

As we will see in the following subsections, particular choices of the basis $\{\ket\alpha\}$ (and the corresponding perturbation Hamiltonians $H'$) lead to F\"orster and modified Redfield theories. Already here, within the general scheme, we see that master equation (\ref{EqMaster}) describes only the dynamics of populations $p_\alpha(t)$, or, in other words, of the diagonal part of the reduced density operator of the system $\sigma(t)=\Tr_B\rho(t)$ in the basis $\{\ket\alpha\}$. The off-diagonal part of it (coherences between different states) are dropped. The calculation of coherences in the framework of this scheme is the subject of the present paper.

Finally, let us note that the choice
\begin{equation}
\mathcal PA=(\Tr_BA)\rho_g,\quad H'=H^{\text{el-ph}}
\end{equation}
followed by an analogous derivation\cite{MayKuhn,QEffBio,Valkunas,Huelga,Davies}  leads to  standard Redfield (or weak-coupling limit) theory and a Markovian quantum master equation for the whole reduced density matrix of the system $\sigma(t)$.

\subsection{F\"orster theory}\label{SecForst}

The F\"orster theory corresponds to the case where the intersite coupling constants $J_{nm}$ are small. We choose $\{\ket\alpha\}=\{\ket n\}$, i.e., the local electronic basis. Correspondingly, $H'=H^{\rm Coul}$, $H_n=\braket{n|H|n}=\varepsilon_n^0+\lambda_n+H^{\rm ph}+u_n$. The application of the general formula (\ref{Eqk}) to this particular case leads to 
\begin{equation}
K_{nm}=2|J_{nm}|^2\Re\int_0^\infty d\tau\,e^{iH_m\tau}e^{-iH_n\tau}\rho_m,
\end{equation}
where $\rho_m=e^{-\beta H_m}/\Tr e^{-\beta H_m}$.
Using the cumulant expansion method (which is approximate in general, but is exact if the reservoir consists of harmonic oscillators,\cite{Mukamel} which is our case), we arrive at:\cite{YangFl}
\begin{equation}\label{EqkF}
K_{nm}=2|J_{nm}|^2\Re\int_0^\infty d\tau\,
F_m^*(\tau)A_n(\tau),
\end{equation}
\begin{eqnarray}
F_m(\tau)&
=&\exp\{-i(\varepsilon_m^0+\lambda_m)\tau\}\nonumber\\
&\times&\Tr \exp\{-i(H_m^{\rm ph}+u_m)\tau\}
\exp\{iH_m^{\rm ph}\tau\}\rho_m^{\rm ph,e}\nonumber\\
&=&\exp\{-i(\varepsilon_m^0-\lambda_m)\tau-g_m^*(\tau)\},
\label{EqFm}
\end{eqnarray}
\begin{eqnarray}
A_n(\tau)&
=&\exp\{-i(\varepsilon_n^0+\lambda_n)\tau\}\nonumber\\
&\times&\Tr \exp\{iH_n^{\rm ph}\tau\}
\exp\{-i(H_n^{\rm ph}+u_n)\tau\}
\rho_n^{\rm ph,g}\nonumber\\
&=&\exp\{-i(\varepsilon_n^0+\lambda_n)\tau-g_n(\tau)\},
\label{EqAn}
\end{eqnarray}
where
\begin{subequations}
\begin{eqnarray}
\rho_n^{\rm ph,g}&&=\frac{\exp\{-\beta H_n^{\rm ph}\}}
{\Tr\exp\{-\beta H_n^{\rm ph}\}},\\
\rho_m^{\rm ph,e}&&=\frac{\exp\{-\beta(H_m^{\rm ph}+u_m)\}}
{\Tr\exp\{-\beta(H_m^{\rm ph}+u_m)\}},
\end{eqnarray}
\end{subequations}
and 
\begin{equation}
g_n(\tau)=\int_0^\tau d\tau_1\int_0^{\tau_1}d\tau_2\,
C_n(\tau_2)
\end{equation}
is the lineshape function,\cite{Mukamel} 
\begin{equation}\label{EqCorr}
C_n(t)=\Tr\{u_n(t)u_n\rho_n^{\rm ph,g}\}
\end{equation}
is the correlation function of the $n$th bath,
$u_n(t)=e^{iH^{\rm ph}_nt}u_ne^{-iH^{\rm ph}_nt}$.

\subsection{Modified Redfield theory}\label{SecMRedf}

Denote
$E_k$ and $\ket k=\sum_{n=1}^N\varphi_{kn}\ket n$  
 the eigenvalues and the corresponding eigenstates of the system Hamiltonian $H^{\rm el}+H^{\rm reorg}+H^{\rm Coul}$ (the spectrum is assumed to be non-degenerate). Taking $\{\ket\alpha\}=\{\ket k\}$, i.e., the  exciton basis, and
\begin{equation}
H'=\sum_{k\neq k'}\ket kH_{kk'}^{\text{el-ph}}\bra{k'},
\end{equation}
where 
\begin{equation}
H_{kk'}^{\text{el-ph}}=\braket{k|H^{\text{el-ph}}|k'}=
\sum_{n=1}^Na_{kk'}(n)u_n,
\end{equation}
$a_{kk'}(n)=\overline{\varphi_{kn}}\varphi_{k'n}$, corresponds to  modified Redfield theory. In other words, this corresponds to the case when the coupling between different delocalized excitations can be treated as small. The application of the general formula (\ref{Eqk}) and the cumulant expansion method to this particular case leads to\cite{YangFl} 
\begin{eqnarray}
K_{kk'}&=&2\Re\int_0^\infty d\tau\,
e^{iH_{k'}\tau}H_{k'k}^{\text{el-ph}}
e^{-iH_k\tau}H_{kk'}^{\text{el-ph}}
\rho_{k'}\nonumber\\
&=&2\Re\int_0^\infty d\tau\,
F_{k'}^*(\tau)A_k(\tau)N_{kk'}(\tau),\label{EqmRedfRate}
\end{eqnarray}
where $\rho_{k}=e^{-\beta H_{k}}/\Tr e^{-\beta H_{k}}$,  $H_k=\braket{k|H|k}$,
\begin{eqnarray}
F_{k'}(\tau)
&=&\exp\{-i(E_{k'}^0-\lambda_{k'})\tau-g_{k'}^*(\tau)\},\\
A_k(t)
&=&\exp\{-i(E_k^0+\lambda_k)\tau-g_k(\tau)\},
\end{eqnarray}
\begin{subequations}
\begin{eqnarray}
N_{kk'}(\tau)&=&\ddot g_{k'kkk'}(\tau)
\exp\{2g_{kkk'k'}(\tau)+2i\lambda_{kkk'k'}\tau\}\nonumber\\
&-&N^{(1)}_{kk'}(\tau)N^{(2)}_{kk'}(\tau),
\end{eqnarray}
\begin{eqnarray}
N^{(1)}_{kk'}(\tau)&=&
[\dot g_{k'kkk}(\tau)-\dot g_{k'kk'k'}(\tau)-2i\lambda_{k'kk'k'}]
\nonumber\\&\times&
\exp\{2g_{kkk'k'}(\tau)+2i\lambda_{kkk'k'}\tau\},\\
N^{(2)}_{kk'}(\tau)&=&
[\dot g_{kk'kk}(\tau)-\dot g_{kk'k'k'}(\tau)-2i\lambda_{kk'k'k'}]
\nonumber\\&\times&
\exp\{2g_{kkk'k'}(\tau)+2i\lambda_{kkk'k'}'\tau\}.
\end{eqnarray}
\end{subequations}
\begin{eqnarray}
g_{k_1k_2k_3k_4}(\tau)&=&
\sum_{n=1}^N a_{k_1k_2}(n)a_{k_3k_4}(n)g_n(\tau),\\
\lambda_{k_1k_2k_3k_4}&=&\sum_{n=1}^N a_{k_1k_2}(n)a_{k_3k_4}(n)\lambda_n,\\
g_k(\tau)&=&g_{kkkk}(\tau),\quad\lambda_k=\lambda_{kkkk}, 
\end{eqnarray}
and $E_k^0=E_k-\lambda_k$.

\section{Dynamics of coherences}\label{SecCoh}

\subsection{The case of equilibrium initial state of the bath}\label{SecEq}

Quantum master equation allows to describe the dynamics of the ``slow'' degrees of freedom $\mathcal P\rho(t)$. In the case of the F\"orster and modified Redfield approaches, this corresponds to the populations $p_n(t)$ and $p_{k}(t)$, respectively. However, the Zwanzig projection operator formalism, namely, formula (\ref{EqQ}) allows also to describe the dynamics of the ``fast'' degrees of freedom as well. A problem is that $\mathcal Q\rho(t)$ is infinite-dimensional, so, expression (\ref{EqQ}) cannot be calculated directly. However, we may be interesting in a particular observable $A$. In this paper we are interested in coherences, i.e., the off-diagonal elements of the density matrix, and we take 
\begin{equation}\label{EqACoh}
A=\ket\alpha\bra\beta,\quad \alpha\neq\beta.
\end{equation}
Denote $A(t)=e^{iH_0t}Ae^{-iH_0t}$, then 
\begin{equation}
\Tr A(t)\rho(t)=
\braket{\beta|\Tr_B\rho(t)|\alpha}=
\sigma_{\beta\alpha}(t),
\end{equation}
where we have denoted $\sigma(t)=\Tr_B\rho(t)$ the reduced density operator of the system.
In this subsection we consider the case $\mathcal Q\rho(0)=0$, i.e., the bath is initially in equilibrium (see Eq.~(\ref{EqRhoEq})). We have
\begin{eqnarray}
\sigma_{\beta\alpha}(t)&=&\Tr A(t)\mathcal P\rho(t)+\Tr A(t)\mathcal Q\rho(t)\label{EqAavgen}
\\
&=&-i\int_0^td\tau\Tr A(t)\mathcal L(\tau)\mathcal P\rho(\tau)
\nonumber\\
&=&-i\int_0^td\tau\Tr A(t)\mathcal L(t-\tau)\mathcal P\rho(t-\tau),
\label{EqAav}
\end{eqnarray}
where we have used Eq.~(\ref{EqQ1}) and 
$\Tr A\mathcal P\rho=\braket{\beta|\mathcal P\rho|\alpha}=0$.  For other choices of $A$ (different from Eq.~(\ref{EqACoh})), the term $\Tr A\mathcal P\rho$ may be non-zero, but it can be treated as known since  $\mathcal P\rho(t)$ is given by the solution of  quantum master equation (\ref{EqNonMark}) or (\ref{EqMark}).

Here we assume that the Born--Markov approximation, which allows to turn Eq.~(\ref{EqNonMark}) into Eq.~(\ref{EqMark}), is valid. Then, again, the time dependence of the term $A(t)\mathcal L(t-\tau)$ in Eq.~(\ref{EqAav}) represents fast bath relaxation processes, while $\mathcal P\rho(t)$ is evolved more slowly. So, we can replace $\mathcal P\rho(t-\tau)$ by $\mathcal P\rho(t)$:
\begin{eqnarray}
\sigma_{\beta\alpha}(t)&=&
-i\int_0^t d\tau\Tr A(t)\mathcal L(t-\tau)
\mathcal P\rho(t)
\nonumber\\
&=&
-i\int_0^t d\tau\Tr A(\tau)\mathcal L
\mathcal P\rho(t)
\nonumber\\
&=&
ip_\beta(t)
\int_0^t d\tau \,\Tr \{e^{iH_\alpha \tau}e^{-iH_\beta \tau}
\rho_\beta H'_{\beta\alpha}\}
\nonumber\\
&-&ip_\alpha(t)
\int_0^t d\tau \,\Tr\{e^{iH_\alpha \tau}e^{-iH_\beta \tau}H'_{\beta\alpha}
\rho_\alpha\}.\label{EqCoh}
\end{eqnarray}
Since $p_\alpha(t)$ and $p_\beta(t)$ are known from the solution of the master equation, formula (\ref{EqCoh}) allows to calculate the dynamics of the coherences.

\begin{remark}\label{RemStages}
Note that, in contrast to Eq.~(\ref{EqMark}), here we do not extend the upper limit of integration to infinity. This approximation works well for large $t$, but introduce a significant error for small $t$ and, in some cases, moderate $t$. Indeed, from Eq.~(\ref{EqCoh}) we see that $\sigma_{\beta\alpha}(0)=0$, which is in agreement with the initial condition $\mathcal Q\rho(0)=0$. However, the extension of the upper limit of integration to infinity would give $\sigma_{\beta\alpha}(0)\neq0$. In contrast, the extension of the upper limit of integration to infinity in Eq.~(\ref{EqNonMark}) produce the error in the time derivative of $\mathcal P\rho(t)$, but not in  $\mathcal P\rho(t)$ directly. Since  $\mathcal P\rho(t)$ is changed slowly, the error in  $\mathcal P\dot\rho(t)$ for small $t$ does not produce a significant error in $\mathcal P\rho(t)$.  Thus, on small times, the coherences evolve faster  than the populations. Also one can notice that the discussed error has the second order of smallness (with respect to $H'$) for $\mathcal P\dot\rho(t)$, but only the first order of smallness for $\sigma_{\beta\alpha}(t)$.

For large times $t$, the integrals in Eq.~(\ref{EqCoh}) saturate to some constant values, so, the time integration can be extended to infinity, and the dynamics of the coherences are driven by the  populations with the constant coefficients:

\begin{equation}\label{EqCohSimp}
\sigma_{\beta\alpha}(t)=c_{\beta\alpha} p_\beta(t)
-d_{\beta\alpha}p_{\alpha}(t),
\end{equation}
where
\begin{subequations}\label{EqCohSimpCoefs}
\begin{eqnarray}
c_{\beta\alpha}&=&i\int_0^\infty
d\tau \,
\Tr \{e^{iH_\alpha \tau}e^{-iH_\beta \tau}
\rho_\beta H'_{\beta\alpha}\}\\
d_{\beta\alpha}&=&
i\int_0^\infty d\tau \,
\Tr\{e^{iH_\alpha \tau}e^{-iH_\beta \tau}H'_{\beta\alpha}
\rho_\alpha\}.
\end{eqnarray}
\end{subequations}
However, as we will observe in Sec.~\ref{SecMap}, the coherences may oscillate around the mean values (\ref{EqCohSimp}) for a long time  comparable with the EET time scale. To describe these oscillations, we shall use more general formula (\ref{EqCoh}).

From Eq.~(\ref{EqCohSimp}) we see that the coherences in F\"orster and modified Redfield theories do not decay to zero (like in the aforementioned phenomenological approach of Ref.~\onlinecite{HwangFu}), but tend to some constant values dependent on the stationary populations. The decay of the coherences under the free (unperturbed) dynamics generated by the Hamiltonian $H_0$ is compensated by their ``pumping'' from the populations induced by $H'$.
\end{remark}

Now we apply the general formula (\ref{EqCoh}) to F\"orster and modified Redfield theories using the cumulant expansion technique. In the F\"orster case  $A=\ket n\bra m$, we obtain
\begin{eqnarray}\label{EqCohF}
\sigma_{mn}(t)&=&
ip_m(t)J_{mn}\int_0^t d\tau\,A_n^*(\tau)F_m(\tau)
\nonumber\\
&-&ip_n(t)J_{mn}\int_0^t d\tau\,F_n^*(\tau)A_m(\tau).
\end{eqnarray}
In the modified Redfield case $A=\ket{k}\bra{k'}$, we obtain
\begin{eqnarray}
\sigma_{k'k}(t)=
&-&p_{k'}(t)\int_0^t d\tau\,A_k^*(\tau)F_{k'}(\tau)N^{(2)}_{kk'}(\tau)^*
\nonumber\\
&-&p_{k}(t)\int_0^t d\tau\,F_{k}^*(\tau)A_{k'}(\tau)
N^{(2)}_{k'k}(\tau).\label{EqCohmR}
\end{eqnarray}

\subsection{The case of non-equilibrium initial state of the bath}\label{SecNoneq}

Now we extend the results of the previous subsection to the case of a non-equilibrium initial state of the bath. Let the initial state of the molecular system be given by Eq.~(\ref{EqRhoNoneq}), i.e., there is still no initial coherences, but the bath is in a non-equilibrium state. The bath state $\rho_g$ describes the bath equilibrium state when no electronic excitation is present. Right after the excitation, the bath is still in the state $\rho_g$ (the Condon approximation), which is non-equilibrium for an excited electronic state. Then
\begin{equation}
\mathcal Q\rho(0)=
\sum_\alpha p_\alpha(0)
\ket\alpha\bra\alpha
\otimes
(\rho_g-\rho_\alpha)\neq0
\end{equation}
 and the first-order approximation of Eq.~(\ref{EqQ}) (with respect to $\mathcal L(t)$) is
\begin{equation}\label{EqQ1noneq}
\mathcal Q\rho(t)=\mathcal Q\rho(0)
-i\int_0^t d\tau\mathcal L(\tau) \mathcal P\rho(\tau)
-i\int_0^t d\tau\mathcal{QL}(\tau) \mathcal Q\rho(0).
\end{equation}
The substitution of Eq.~(\ref{EqQ1noneq}) into Eq.~(\ref{EqAavgen}) gives
\begin{eqnarray}
\sigma_{\beta\alpha}(t)=
&-&i\int_0^t d\tau\Tr A(\tau)\mathcal L
\mathcal P\rho(t)
\nonumber\\
&-&i\int_0^t d\tau\Tr A(t)\mathcal L(t-\tau)
\mathcal Q\rho(0)\label{EqAnoneq}
\end{eqnarray}
The first term has been already calculated in Eq.~(\ref{EqCoh}). Consider the second term:
\begin{eqnarray}
&&\Tr A(t)\mathcal L(t-\tau)
\mathcal Q\rho(0)
\nonumber\\&=&
\Tr A(\tau)\mathcal L
[e^{-iH_0(t-\tau)}\mathcal Q\rho(0)e^{iH_0(t-\tau)}]
\nonumber\\&=&
p_\alpha(0)\Tr e^{iH_\alpha\tau}e^{-iH_\beta\tau}
H'_{\beta\alpha}
[
\rho_g^\alpha(t-\tau)-\rho_\alpha
]
\nonumber\\&-&
p_\beta(0)\Tr e^{iH_\alpha\tau}e^{-iH_\beta\tau}
[
\rho_g^\beta(t-\tau)-\rho_\beta
]
H'_{\beta\alpha},\label{EqDopTerm}
\end{eqnarray}
where $\rho_g^\alpha(t)=e^{-iH_\alpha t}\rho_ge^{iH_\alpha t}$. So, $\sigma_{\beta\alpha}(t)$ has the form
\begin{equation}\label{EqCohIneq}
\sigma_{\beta\alpha}(t)=\sigma_{\beta\alpha}^{\rm eq}(t)+
\sigma_{\beta\alpha}^{\rm noneq}(t),
\end{equation}
where $\sigma_{\beta\alpha}^{\rm eq}(t)$ is the expression for the equilibrium case given by Eq.~(\ref{EqCoh}) and 
$\sigma_{\beta\alpha}^{\rm noneq}(t)$ is a non-equilibrium correction:

\begin{eqnarray}
\sigma_{\beta\alpha}^{\rm noneq}(t)&&=
ip_\alpha(0)
\int_0^t d\tau \,\Tr e^{iH_\alpha \tau}e^{-iH_\beta \tau}H'_{\beta\alpha}
\rho_\alpha
\nonumber\\
&&-ip_\beta(0)
\int_0^t d\tau \,\Tr e^{iH_\alpha \tau}e^{-iH_\beta \tau}
\rho_\beta H'_{\beta\alpha}
\nonumber\\
-ip_\alpha(0)&&\int_0^t d\tau \,
\Tr e^{iH_\alpha t}e^{-iH_\beta\tau}
H'_{\beta\alpha}e^{-iH_\alpha(t-\tau)}\rho_g
\nonumber\\
+ip_\beta(0)&&\int_0^t d\tau \,
\Tr e^{iH_\alpha\tau}e^{-iH_\beta t}
\rho_ge^{iH_\beta(t-\tau)}H'_{\beta\alpha}.\label{EqCohIneqAdd}
\end{eqnarray}
The non-equilibrium terms are significantly non-zero only on the time scale of the bath relaxation. Since the populations evolve more slowly, we can replace $p_\alpha(0)$ and $p_\beta(0)$ by $p_\alpha(t)$ and $p_\beta(t)$ in Eq.~(\ref{EqCohIneqAdd}) and obtain
\begin{equation}\label{EqCoherIneq2}
\begin{split}
\sigma_{\beta\alpha}(t)&=
ip_\beta(t)\int_0^t d\tau \,
\Tr e^{iH_\alpha\tau}e^{-iH_\beta t}
\rho_ge^{iH_\beta(t-\tau)}H'_{\beta\alpha}
\\
&-ip_\alpha(t)\int_0^t d\tau \,
\Tr e^{iH_\alpha t}e^{-iH_\beta\tau}
H'_{\beta\alpha}e^{-iH_\alpha(t-\tau)}\rho_g.
\end{split}
\end{equation}
However, in the calculations below, this additional approximation introduces small but noticeable error and, so, we will use formulas (\ref{EqCohIneq})--(\ref{EqCohIneqAdd}).

\begin{remark}
The substitution of  Eq.~(\ref{EqQ1noneq}) into Eq.~(\ref{EqPrho}) yields Eq.~(\ref{EqMark}) with a non-equilibrium correction, which is calculated in Ref.~\onlinecite{Seibt}. Here, for simplicity, we assume that the bath relaxes much faster than the population transfer occurs and this correction is negligible. Then $p_\alpha(t)$ and $p_\beta(t)$ are well described by master equation (\ref{EqMaster}) with the equilibrium transfer rates (\ref{Eqk}).

But even in this case, the non-equilibrium corrections for coherences (\ref{EqCohIneqAdd})  may be essential because, on small times, coherences evolve faster than the populations, see Remark~\ref{RemStages}. But in general,  solutions of the master equation with the non-equilibrium (time-dependent) transfer rates from Ref.~\onlinecite{Seibt} also can be used for $p_\alpha(t)$ and $p_\beta(t)$  in Eqs.~(\ref{EqCoh}), (\ref{EqCohIneq}), and (\ref{EqCoherIneq2}). Also note that the non-equilibrium generalization of the F\"orster transfer rates was firstly derived in Ref.~\onlinecite{ForsterNoneq}, but a compact expression involving the lineshape functions $g_n(t)$ was obtained in Ref.~\onlinecite{Seibt}.
\end{remark}

The traces involving $\rho_g$ in Eqs.~(\ref{EqCohIneqAdd}) and (\ref{EqCoherIneq2}) for F\"orster and modified Redfield theories appear also during the derivation of the non-equilibrium transfer rates for the populations and were also calculated in Ref.~\onlinecite{Seibt}. For F\"orster theory, Eq.~(\ref{EqCohIneqAdd})  takes the form

\begin{equation}\label{EqCohIneqF}
\begin{split}
\sigma_{mn}^{\rm noneq}(t)&=
ip_m(0)J_{mn}
\int_0^t d\tau \,
A_n^*(\tau)[F_m(\tau,t)-F_m(\tau)]
\\
&-ip_n(0)J_{mn}
\int_0^t d\tau \,
[F_n(\tau,t)-F_n(\tau)]^*A_m(\tau),
\end{split}
\end{equation}
where 
\begin{eqnarray}
F_m(\tau,t)=
\exp\{-&&i(\varepsilon_m^0+\lambda_m)\tau-g_m^*(\tau)
\nonumber\\
-&&2i\Im[g_m(t)-g_m(t-\tau)]
\}.\label{EqFmt}
\end{eqnarray}
Note that $F_m(\tau,t)$ can be replaced by $F_m(\tau)$ in Eq.~(\ref{EqCohIneqF}) for large $t$ because  $\lim_{\tau\to\infty}\dot g_m(\tau)=-\lambda_m$\cite{Zhang} and the integral expressions are significantly non-zero only for small $\tau$. So, the non-equilibrium correction vanishes for large times.

For  modified Redfield theory, Eq.~(\ref{EqCohIneqAdd})  takes the form
\begin{eqnarray}
\sigma_{k'k}^{\rm noneq}(t)
=
-p_{k'}(0)
\int_0^t d\tau 
&&A_k^*(\tau)\big[F_{k'}(\tau,t)N^{(2)}_{kk'}(\tau,t)^*\nonumber\\
-
&&F_{k'}(\tau)N^{(2)}_{kk'}(\tau)^*\big]\nonumber
\\
-p_k(0)
\int_0^t d\tau \,
\big[&&F_k(\tau,t)^*N^{(2)}_{k'k}(\tau,t)\nonumber\\
-&&F_k(\tau)^*N^{(2)}_{k'k}(\tau)\big]
A_{k'}(\tau),\label{EqCohIneqR}
\end{eqnarray}
where 
\begin{eqnarray}
F_k(\tau,t)=
\exp\{-&&i(\varepsilon_k^0+\lambda_k)\tau-g_k^*(\tau)
\nonumber\\
-&&2i\Im[g_k(t)-g_k(t-\tau)]
\},
\end{eqnarray}
\begin{multline}
N^{(2)}_{kk'}(\tau,t)\\=
[\dot g_{kk'kk}(\tau)-\dot g_{kk'k'k'}(\tau)
+2i(\Im\dot g)_{kk'k'k'}(t-\tau)]
\\\times
\exp\{2g_{kkk'k'}(\tau)
-2i\Im[g_{kkk'k'}(t)-g_{kkk'k'}(t-\tau)]\},
\end{multline}
\begin{equation}
(\Im\dot g)_{kk'k'k'}(t)=\sum_{n=1}^Na_{kk'}(n)a_{k'k'}(n)\Im\dot g_n(t).
\end{equation}
Note that $(\Im\dot g)_{kk'k'k'}(t)=\Im\dot g_{kk'k'k'}(t)$ whenever $a_{kk'}(n)$   is real for all $n$.

\subsection{The case of initial coherences in the system}\label{SecIni}

Finally, we consider the case when the initial state has non-zero off-diagonal elements. We will consider the initial state of the form
\begin{equation}\label{EqRhoIni}
\rho(0)=\sigma(0)\otimes\rho_g,
\end{equation} 
where $\sigma(0)$ is an arbitrary density operator of the electronic degrees of freedom. We divide $\mathcal Q\rho(0)$ into the diagonal and off-diagonal parts:
\begin{subequations}\label{EqQdiagoff}
\begin{equation}
\mathcal Q\rho(0)=\mathcal Q\rho(0)^{\rm diag}+
\mathcal Q\rho(0)^{\text{off-diag}},
\end{equation}
\begin{eqnarray}
\mathcal Q\rho(0)^{\rm diag}&=&
\sum_\alpha p_{\alpha}(0)
\ket\alpha\bra\alpha
\otimes
(\rho_g-\rho_\alpha),
\\
\mathcal Q\rho(0)^{\text{off-diag}}&=&
\sum_{\beta\neq\alpha}\sigma_{\beta\alpha}(0)
\ket\beta\bra\alpha
\otimes
\rho_g.
\end{eqnarray}
\end{subequations}
In both Eqs.~(\ref{EqPrho}) and (\ref{EqAavgen}) only the off-diagonal part of  $\mathcal Q\rho(t)$ matters. We have

\begin{eqnarray}
\mathcal Q\rho(t)^{\text{off-diag}}&=&
\mathcal Q\rho(0)^{\text{off-diag}}
-i\int_0^t d\tau\mathcal L(\tau) \mathcal P\rho(\tau)
\nonumber\\
&-&i\int_0^t d\tau\mathcal{L}(\tau) \mathcal Q\rho(0)^{\rm diag}.
\label{EqQ1noneqoffd}
\end{eqnarray} 
Note that the influence of $\mathcal Q\rho(0)^{\rm diag}$ (of the last term in Eq.~(\ref{EqQ1noneqoffd})) has been already taken into account in the previous subsection. Here we will study the influence of only the off-diagonal part  $\mathcal Q\rho(0)^{\text{off-diag}}$.

Firstly, the initial coherence infer the population transfer on small times. As we discussed above, the substitution of the last term of Eq.~(\ref{EqQ1noneqoffd}) into Eq.~(\ref{EqPrho}) produces the non-equilibrium correction to the population transfer rates which is assumed to be negligible. The substitution of the first two terms of the right-hand side of Eq.~(\ref{EqQ1noneqoffd}) into Eq.~(\ref{EqPrho}) yields Eq.~(\ref{EqMark}) with the following additional term in the right-hand side:

\begin{equation}
\begin{split}
&-i\mathcal P\mathcal L(t)\mathcal Q\rho(0)^{\text{off-diag}}\\
&=
2\Im\sum_{\alpha<\beta}
\sigma_{\alpha\beta}(0)
\Tr\{H'_{\beta\alpha}(t)\rho_g\}
(\ket{\beta}\bra{\beta}\rho_{\beta}-
\ket{\alpha}\bra{\alpha}\rho_{\alpha}),
\end{split}
\end{equation}
where  
\begin{equation}
H'_{\beta'\alpha}(t)=
\exp\{iH_{\beta'}t\}H'_{\beta'\alpha}\exp\{-iH_{\alpha}t\}.
\end{equation} 
Consequently, the master equation (\ref{EqMaster}) is modified to
\begin{eqnarray}
\dot p_\alpha(t)=\sum_{\beta\neq\alpha}
\big(&&K_{\alpha\beta}p_\beta(t)-K_{\beta\alpha}p_\alpha(t)
\nonumber\\
-&&2\Im\{\sigma_{\alpha\beta}(0)\Tr[H'_{\beta\alpha}(t)\rho_g]\}\big),
\label{EqMasterIni}
\end{eqnarray}

Now let us consider the influence of initial coherences on their further dynamics.  The substitution of  Eq.~(\ref{EqQ1noneqoffd}) into Eq.~(\ref{EqAavgen}) gives
$\sigma_{\beta\alpha}(t)$ as a sum of three terms:
\begin{equation}\label{EqCohGen}
\sigma_{\beta\alpha}(t)=
\sigma_{\beta\alpha}^{\rm eq}(t)+\sigma_{\beta\alpha}^{\rm noneq}(t)
+\sigma_{\beta\alpha}^{\rm init}(t),
\end{equation}
where the basic term $\sigma_{\beta\alpha}^{\rm eq}(t)$ is given by Eq.~(\ref{EqCoh}), the non-equilibrium correction $\sigma_{\beta\alpha}^{\rm noneq}(t)$ is given by Eq.~(\ref{EqCohIneqAdd}), and the new term (a correction caused by the initial coherence) $\sigma_{\beta\alpha}^{\rm init}(t)$ is given by
\begin{equation}\label{EqCohIni}
\begin{split}
&\sigma_{\beta\alpha}^{\rm init}(t)
=\Tr A(t)\mathcal Q\rho(0)^{\text{off-diag}}\\
&-i\int_0^t d\tau\Tr A(t)\mathcal L(t-\tau)
\mathcal Q\rho(0)^{\text{off-diag}}\\
&=
\sigma_{\beta\alpha}(0)
\Tr\{e^{iH_\alpha t}e^{-iH_\beta t}\rho_g\}
\\
&+i\sum_{\gamma\neq\alpha,\beta}\sigma_{\beta\gamma}(0)
\int_0^td\tau
\Tr\{e^{iH_\gamma(t-\tau)}H'_{\gamma\alpha}
e^{iH_\alpha\tau}e^{-iH_\beta t}\rho_g\}\\
&-i\sum_{\gamma\neq\alpha,\beta}\sigma_{\gamma\alpha}(0)\!
\int_0^td\tau
\Tr\{e^{iH_\alpha t}e^{-iH_\beta\tau}H'_{\beta\gamma}
e^{-iH_\gamma(t-\tau)}\!
\rho_g\}.
\end{split}
\end{equation}

Now let us apply Eqs.~(\ref{EqMasterIni}) and (\ref{EqCohIni}) to the particular theories. In F\"orster theory, the traces in Eqs.~(\ref{EqMasterIni}) and (\ref{EqCohIni}) can be easily calculated, which results in
\begin{multline}\label{EqMasterIniF}
\dot p_n(t)=\sum_{m\neq n}
\big(K_{nm}p_m(t)-K_{mn}p_n(t)
\\
-2\Im\{\sigma_{nm}(0)J_{mn}A_n(t)A^*_m(t)\}\big),
\end{multline}
and
\begin{eqnarray}
\sigma_{mn}^{\rm init}(t)
=\sigma_{mn}(0)&&A_n^*(t)A_m(t)
\nonumber\\
+\,i\sum_{l\neq n,m}
\int_0^td\tau\,
\{&&\sigma_{ml}(0)A_n^*(\tau)A_m(t)A_l^*(t-\tau)
\nonumber
\\
-&&\sigma_{ln}(0)A_n^*(t)A_m(\tau)A_l(t-\tau)\}.\label{EqCohIniF}
\end{eqnarray}

In the modified Redfield theory, the trace in Eq.~(\ref{EqMasterIni}) is a particular case of the traces calculated in Ref.~\onlinecite{Seibt}, while the trace in Eq.~(\ref{EqCohIni}) can be calculated by a slight modification of the derivations there. It turns out that
\begin{multline}
\dot p_k(t)=\sum_{k'\neq k}
\big(K_{kk'}p_{k'}(t)-K_{k'k}p_k(t)
\\
+2\Re\big\{\sigma_{kk'}(0)A_k(t)A^*_{k'}(t)
\exp\{2\Re g_{kkk'k'}(t)\}
\\
\times
[\dot g_{k'kkk}(t)-\dot g^*_{kk'k'k'}(t)]
\big\}\big),
\label{EqMasterIniR}
\end{multline}
where 
\begin{multline}
\sigma_{k'k}^{\rm init}(t)=
\sigma_{k'k}(0)A_k^*(t)A_{k'}(t)\exp\{2\Re g_{kkk'k'}(t)\}
\\
+\sum_{k''\neq k,k'}\!
\int_0^td\tau
\{\sigma_{k''k}(0)A_k^*(t)A_{k'}(\tau)A_{k''}(t-\tau)
N^{\rm (init)}_{kk'k''}(\tau,t)\\
+\sigma_{k'k''}(0)A_k^*(\tau)A_{k'}(t)A_{k''}^*(t-\tau)
N^{\rm (init)}_{k'kk''}(\tau,t)^*\},
\label{EqCohIniR}
\end{multline}
where
\begin{eqnarray}
&&N^{\rm (init)}_{kk'k''}(\tau,t)\nonumber
\\&&\quad=\exp\{g_{kkk'k'}^*(t)-g_{kkk'k'}^*(t-\tau)+g_{kkk'k'}(\tau)\}
\nonumber\\
&&\quad\times\exp\{g_{k'k'k''k''}(t-\tau)-g_{k'k'k''k''}(t)+g_{k'k'k''k''}(\tau)\}
\nonumber\\
&&\quad\times\exp\{g_{kkk''k''}(t)+g_{kkk''k''}^*(t-\tau)-g_{kkk''k''}(\tau)\}
\nonumber\\
&&\quad\times[\dot g_{kkk'k''}(\tau)-\dot g_{k'k'k'k''}(\tau)+\dot g_{kkk''k'}^*(t-\tau)
\nonumber\\
&&\quad\:-\dot g_{k'k''k''k''}(t-\tau)].
\end{eqnarray}

\begin{remark}
The following question  about self-consistency of our calculations may be asked. Let $\mathcal Q\rho(0)=0$. At some instant $t_0>0$, coherences are non-zero and given by Eq.~(\ref{EqCoh}). If we apply the calculations of this subsection for the case of initial coherences, do we reproduce formulas (\ref{EqMaster}), (\ref{Eqk}), and (\ref{EqCoh}) for an arbitrary time $t>t_0$? An important point we should notice is that $\rho(t_0)$ differs from the state given by Eq.~(\ref{EqRhoIni}). Namely,
\begin{subequations}\label{EqRhot0}
\begin{equation}
\mathcal Q\rho(t_0)=-i\int_0^{t_0}\!d\tau
\mathcal L(\tau)\mathcal P\rho(\tau)=
\sum_{\alpha\neq\beta}\ket\beta
\braket{\beta|\mathcal Q\rho(t)|\alpha}\bra\alpha,
\end{equation}
\begin{multline}
\braket{\beta|\mathcal Q\rho(t)|\alpha}\\
=i\int_0^t d\tau
\{
p_\beta(t)\rho_\beta e^{iH_\beta(t-\tau)}H'_{\beta\alpha}e^{-iH_\alpha}(t-\tau)
\\
-p_\alpha(t)e^{iH_\beta(t-\tau)}H'_{\beta\alpha}
e^{-iH_\alpha(t-\tau)}\rho_\alpha\}.
\end{multline}
\end{subequations}
So, the calculations of this subsection should be modified for $\mathcal Q\rho(t_0)$ given by Eqs.~(\ref{EqRhot0}). If we do this, then, since $\mathcal Q\rho(t)$ can be expressed in two ways as
\begin{eqnarray}
\mathcal Q\rho(t)
&=&
-i\int_0^{t_0}d\tau\,
\mathcal L(\tau)\mathcal P\rho(\tau)
-
i\int_{t_0}^{t}d\tau\,
\mathcal L(\tau)\mathcal P\rho(\tau)\nonumber\\
&=&-i\int_{0}^t d\tau\,
\mathcal L(\tau)\mathcal P\rho(\tau),
\end{eqnarray}
we will reproduce formulas (\ref{EqMaster}), (\ref{Eqk}), and (\ref{EqCoh}). Informally speaking, the coherences and the state of the bath in Eqs.~(\ref{EqRhot0}) are not arbitrary, but adjusted with each other in a special way.
\end{remark}

\section{Calculations}\label{SecCalc}

In this section we will compare the calculations of the coherences according to the derived formulas  with the calculations according to the numerically exact HEOM (hierarchical equations of motion) method.\cite{IFl} The lineshape functions $g_n(t)$ can be expressed as\cite{Mukamel}
\begin{eqnarray}
g_n(t)&=&
\frac1{2\pi}
\int_{-\infty}^{+\infty}
d\omega\,
(1-\cos\omega t)
\coth\left(\frac{\beta\omega}2\right)
\mathcal J_n(\omega)\nonumber\\
&+&\frac i{2\pi}
\int_{-\infty}^{+\infty}
d\omega\,
(\sin\omega t-\omega t)
\mathcal J_n(\omega),\label{EqgJ}
\end{eqnarray}
where $\mathcal J_n(\omega)$ is the spectral density function for the bath coupled to the $n$th site defined as 
\begin{equation}
\mathcal J_n(\omega)=\frac\pi2\sum_iM_{ni}\omega_{ni}d_{ni}^2
[\delta(\omega-\omega_{ni})-\delta(\omega+\omega_{ni})].
\end{equation}
It can be expressed in terms of the correlation function $C_n(t)$ (see Eq.~(\ref{EqCorr})) as
\begin{equation}
\omega^2\mathcal J_n(\omega)=i\int_{-\infty}^{+\infty}dt\,e^{i\omega t}
\Im\left[
C_n(t)\right].
\end{equation}

We will consider the dimer case $N=2$ and assume that the spectral density function for both molecules has the Drude--Lorentz form:
\begin{equation}\label{EqDrude}
\omega^2\mathcal J(\omega)=
2\lambda\frac{\omega\gamma}{\omega^2+\gamma^2},
\end{equation}
where $\lambda$ and $\gamma$ are the reorganization energy and Debye frequency, which are identical for both molecules. The reorganization energy characterizes the strength of the coupling of electronic and phononic degrees of freedom, and the Debye frequency characterizes a time scale of fluctuation of the electronic energy and dissipation of the phonon reorganization energy. 

Also we impose a high-temperature condition $\beta\gamma<1$, where $\beta=1/k_{\rm B}T$ is the inverse temperature of the baths, $T$ is the temperature and $k_{\rm B}$ is the Boltzmann's constant. Then we can approximate $\coth(\beta\omega/2)$ by $2/(\beta\omega)$ in Eq.~(\ref{EqgJ}) and obtain the expression for the lineshape function (the same for both monomers):
\begin{equation}\label{Eqg}
g(t)=\frac\lambda\gamma
\left(\frac2{\beta\gamma}-i\right)
\left(e^{-\gamma t}+\gamma t-1\right).
\end{equation}

In all examples we put $T=300~\rm{K}$ and $\gamma^{-1}=100~\rm{fs}$ ($\gamma=53.08~\rm{cm}^{-1}$). Then $\beta\gamma\approx0.24$. Also in all examples, $\varepsilon^0_1-\varepsilon^0_2=100~\rm{cm}^{-1}$. Since $J_{12}$ is the only intersite dipole coupling, we will denote it as $J$. We will vary the parameters $J$ and $\lambda$.

To distinguish between the matrix elements of the electronic density operator in the local basis $\sigma_{mn}$ and in the exciton basis $\sigma_{k'k}$ when the indexes are concrete numbers, we will use the following denotations: the matrix elements in the local electronic basis will be denoted as $\sigma_{mn}$, while those in the exciton basis will be denoted as $\sigma^{\rm ex}_{k'k}$. The same rule will be applied to the Dirac notations and the bath equilibrium states: $\ket k$ and $\rho_k$ will be redenoted as $\ket k_{\rm ex}$ and $\rho_k^{\rm ex}$.

\subsection{F\"orster theory}

\begin{figure}[h]
\begin{centering}
\includegraphics[width=1\columnwidth]{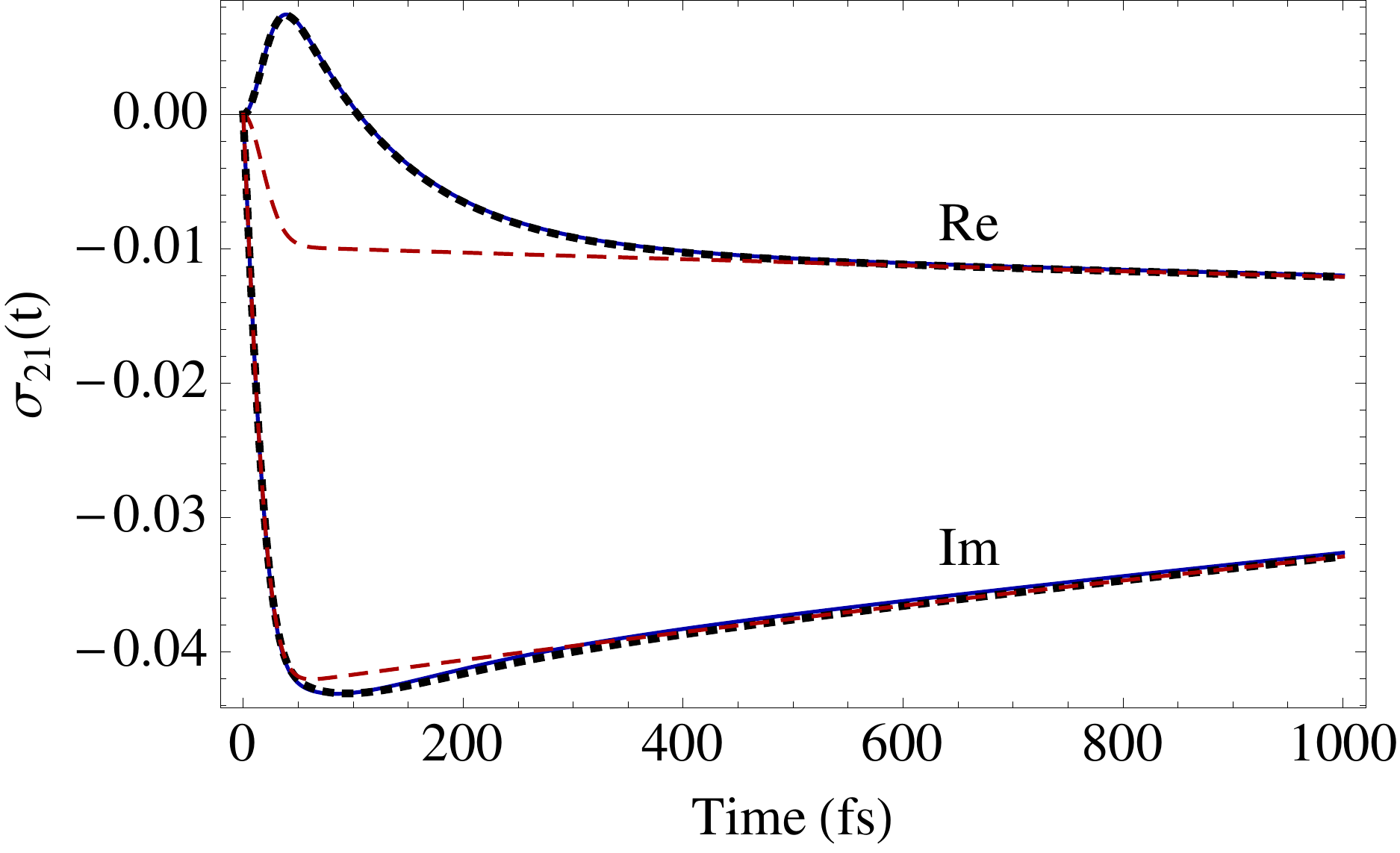}
\end{centering}
\vskip -4mm
\caption
{
Calculations of the electronic coherence $\sigma_{21}(t)$ within F\"orster theory according to formula (\ref{EqCohF}) with and without the non-equilibrium correction (\ref{EqCohIneqF}) in comparison with the numerically exact calculations according to HEOM for the parameters $J=10~\rm{cm}^{-1}$ and $\lambda=100~\rm{cm}^{-1}$, and the initial state (\ref{EqIniket1}). Blue solid lines: HEOM, black thick dotted lines: F\"orster theory with the non-equilibrium correction, red dashed lines: F\"orster theory without the non-equilibrium correction.}
\label{FigCohF}
\end{figure}

Let us consider the parameters 
$J=10~\rm{cm}^{-1}$ and $\lambda=100~\rm{cm}^{-1}$,
and the initial state
\begin{equation}\label{EqIniket1}
\rho(0)=\ket{1}\bra{1}\rho_g
\end{equation}
(a particular case of Eq.~(\ref{EqRhoNoneq})).
The results of the calculations of the electronic coherence $\sigma_{21}(t)$ within F\"orster theory according to the derived formula (\ref{EqCohF}) with and without the non-equilibrium correction (\ref{EqCohIneqF})  in comparison with HEOM is presented on Fig.~\ref{FigCohF}. We see an excellent agreement of the calculations according to Eqs.~(\ref{EqCohF}) and (\ref{EqCohIneqF}) with the numerically exact result. As for Eq.~(\ref{EqCohF}) without the non-equilibrium correction, it gives the same results for large times, but fails to describe the coherence (the real part of it) on small times. 

\begin{figure}[h]
\begin{centering}
\includegraphics[width=1\columnwidth]{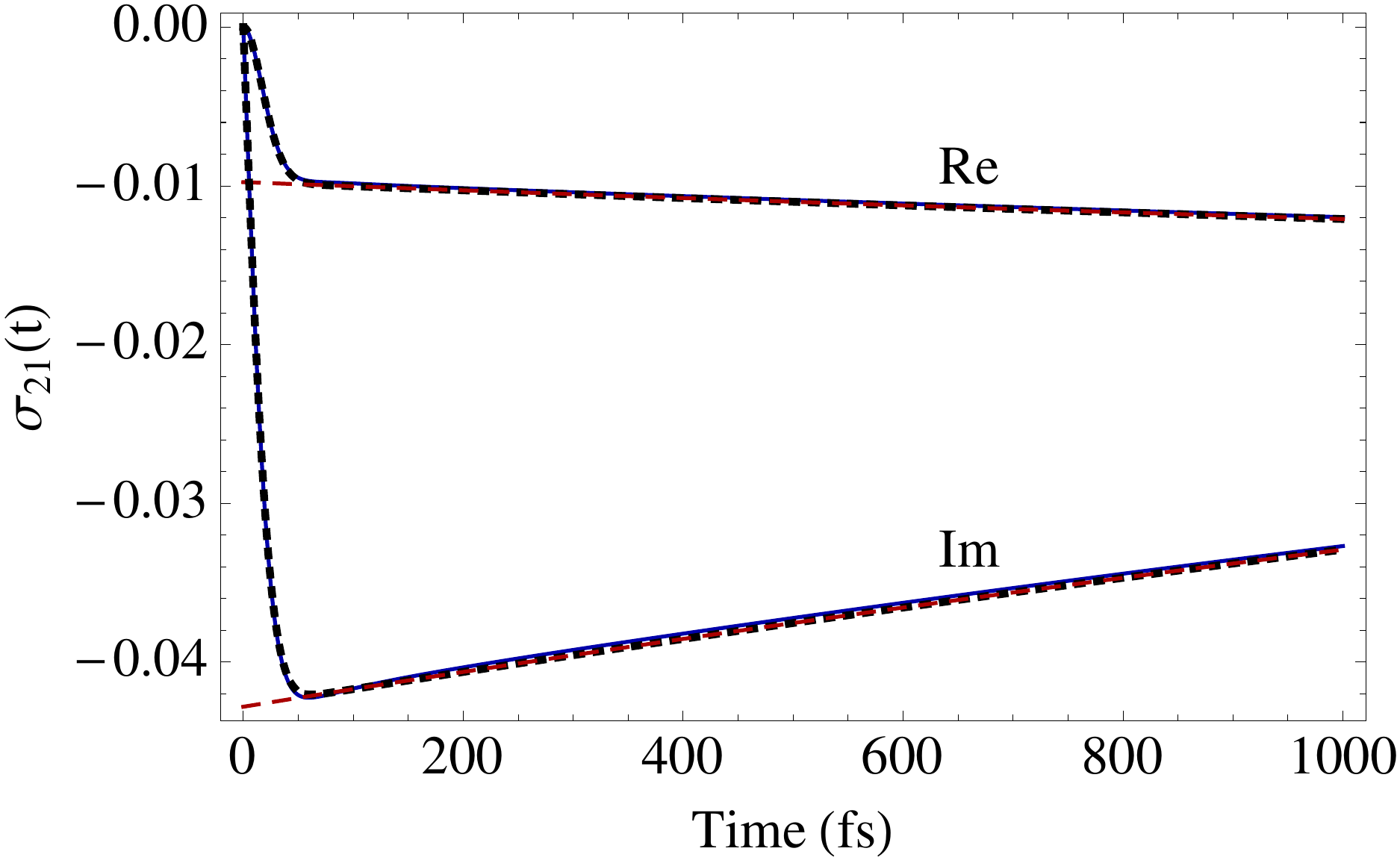}
\end{centering}
\vskip -4mm
\caption
{
The same as Fig.~\ref{FigCohF}, but the initial bath state is equilibrium, Eq.~(\ref{EqIniket1eq}). Blue solid  lines: HEOM, black thick dotted line:  F\"orster theory (without the non-equilibrium correction), red dashed line: simplified (``Markovian'') formula (\ref{EqCohSimp}).}
\label{FigCohEqF}
\end{figure}

\begin{figure}[h]
\begin{centering}
\includegraphics[width=1\columnwidth]{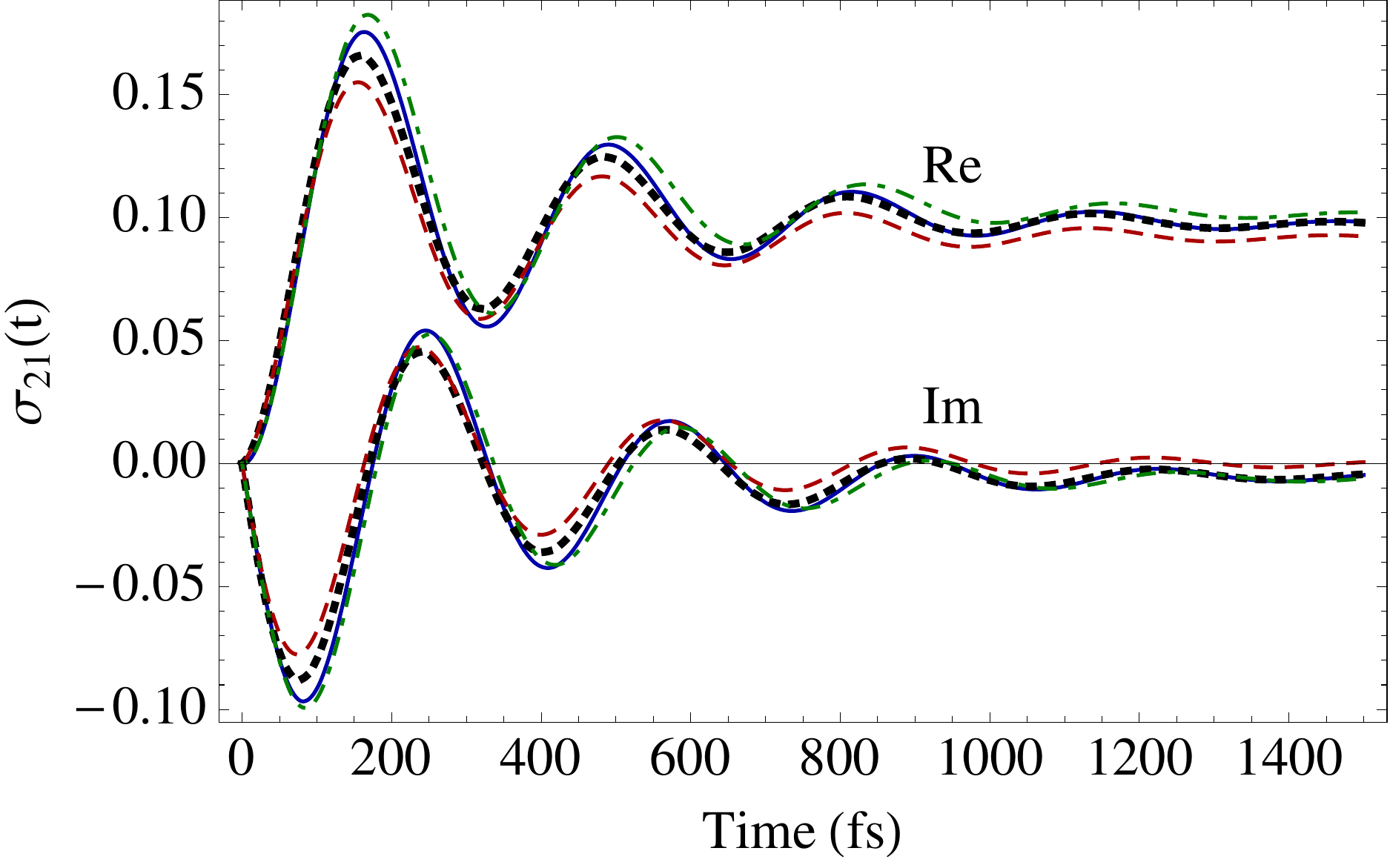}
\end{centering}
\vskip -4mm
\caption
{
Calculations of the electronic coherence $\rho_{21}(t)$ within F\"orster theory (without the non-equilibrium correction) and within modified Redfield theory (Eq.~(\ref{EqMasterIniR}) for exciton populations and Eq.~(\ref{EqCohmR}) with correction (\ref{EqCohIniR}) for exciton coherences) in comparison with HEOM and standard Redfield theory for the parameters $J=10~\rm{cm}^{-1}$ and $\lambda=1~\rm{cm}^{-1}$, and the initial state (\ref{EqIniket1}). Blue solid lines: HEOM and modified Redfield (there are no visible differences between them on this scale), black thick dotted lines: F\"orster theory, red dashed line: the standard Redfield equation without secular approximation, green dot-dashed line: the standard Redfield equation with secular approximation. We see that, for fast nuclear motion, the non-equilibrium correction is unnecessary. Also F\"orster theory provides good results, despite the regime $\lambda\ll J$.}
\label{FigCohEqF2}
\end{figure}

The calculations according to Eq.~(\ref{EqCohF}) without the non-equilibrium correction can be compared with the calculations according to HEOM with the equilibrium initial state
\begin{equation}\label{EqIniket1eq}
\rho(0)=\ket1\bra1\rho_{1}
\end{equation}
(a particular case of the state (\ref{EqRhoEq})) is given on Fig.~\ref{FigCohEqF}. We see an excellent agreement. 
The present choice of parameters corresponds to the case of slow nuclear motion.\cite{MayKuhn,Mukamel} Let us choose the parameters 
corresponding to fast nuclear motion:
$J=10~\rm{cm}^{-1}$, $\lambda=1~\rm{cm}^{-1}$.
Then the non-equilibrium corrections are negligible as shown on Fig.~\ref{FigCohEqF2}. Also we see that F\"orster theory provides good results, despite the regime $\lambda\ll J$, while  F\"orster theory is believed to be valid only in the inverse case $J\ll\lambda$. F\"orster theory gives here good results also for the populations $\sigma_{11}(t)$ and $\sigma_{22}(t)$: the error does not exceed 3.5\% for all times. We discuss the range of validity of  F\"orster theory in Sec.~\ref{SecVal}.

Let us return to the case $J=10~\rm{cm}^{-1}$ and $\lambda=100~\rm{cm}^{-1}$, when the non-equilibrium correction is essential, and consider the case of initial electronic coherences. Namely, let us consider the case when the initial electronic state is not a local (electronic) excitation, but an exciton eigenstate:
\begin{equation}\label{EqIniket1ex}
\rho(0)=\ket{1}{}_{\rm ex}\bra{1}\rho_g.
\end{equation}
Using the derived formulas, we can calculate the dynamics of the whole density matrix (not only its diagonal part in the local basis) in the framework of F\"orster theory. Let us calculate, in the framework of F\"orster theory, the exciton population $\sigma_{11}^{\rm ex}(t)$ as well as the exciton coherence $\sigma_{21}^{\rm ex}(t)$ and compare the results with the calculations within the modified Redfield theory, which is traditionally used  for the description of exciton population transfer. An exciton population is a linear combination of the local site populations and intersite (electronic) coherences. Since we use the non-equilibrium corrections for the electronic coherences, we use also non-equilibrium, time-dependent generalizations for population transfer rates $K_{\alpha\beta}$ in Eq.~(\ref{EqMaster}) for both modified Redfield and F\"orster theories,\cite{Seibt} for the honest comparison between the two theories. This is the only place in our work where we have used the non-equilibrium, time-dependent population transfer rates.

\begin{figure}[h]
\begin{centering}
\includegraphics[width=1\columnwidth]{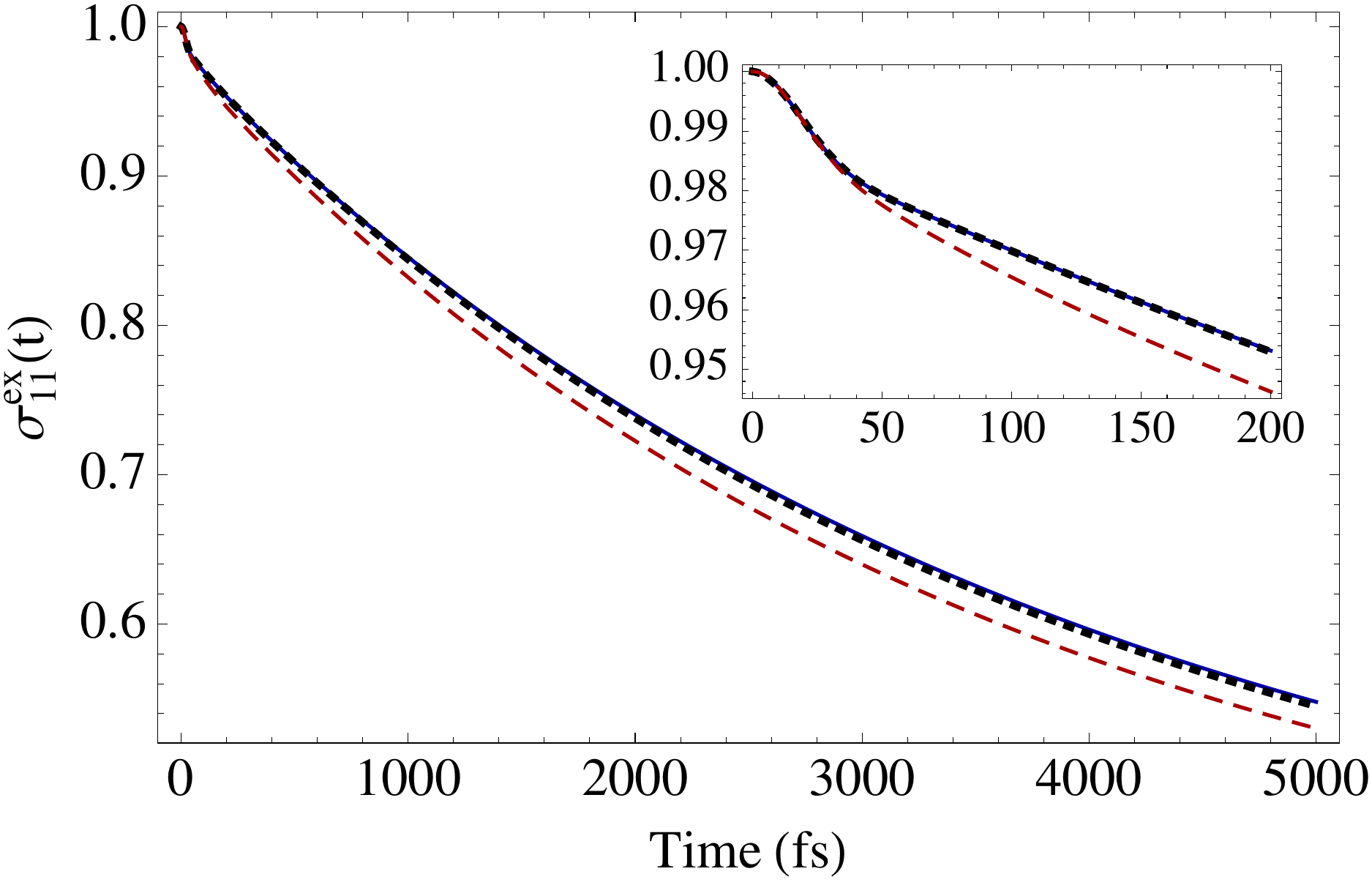}
\end{centering}
\vskip -4mm
\caption
{
Calculations of the exciton population  $\sigma^{\rm ex}_{11}(t)$ within F\"orster theory in comparison with the numerically exact calculations according to HEOM and with modified Redfield theory for the parameters $J=10~\rm{cm}^{-1}$ and $\lambda=100~\rm{cm}^{-1}$, and the initial state (\ref{EqIniket1ex}). Blue solid lines: HEOM, black thick dotted lines: F\"orster theory (Eqs.~(\ref{EqMaster}), (\ref{EqkF}), (\ref{EqMasterIniF}), (\ref{EqCohGen}), (\ref{EqCohF}), (\ref{EqCohIneqF}), and (\ref{EqCohIniF})), red dashed lines: modified Redfield theory (Eqs.~(\ref{EqMaster}) and (\ref{EqmRedfRate})). For both F\"orster and modified Redfield theories, instead of $K_{\alpha\beta}$ given by Eq.~(\ref{Eqk}), we have used the non-equilibrium, time-dependent generalizations.}
\label{FigFini}
\end{figure}

\begin{figure}[h]
\begin{centering}
\includegraphics[width=1\columnwidth]{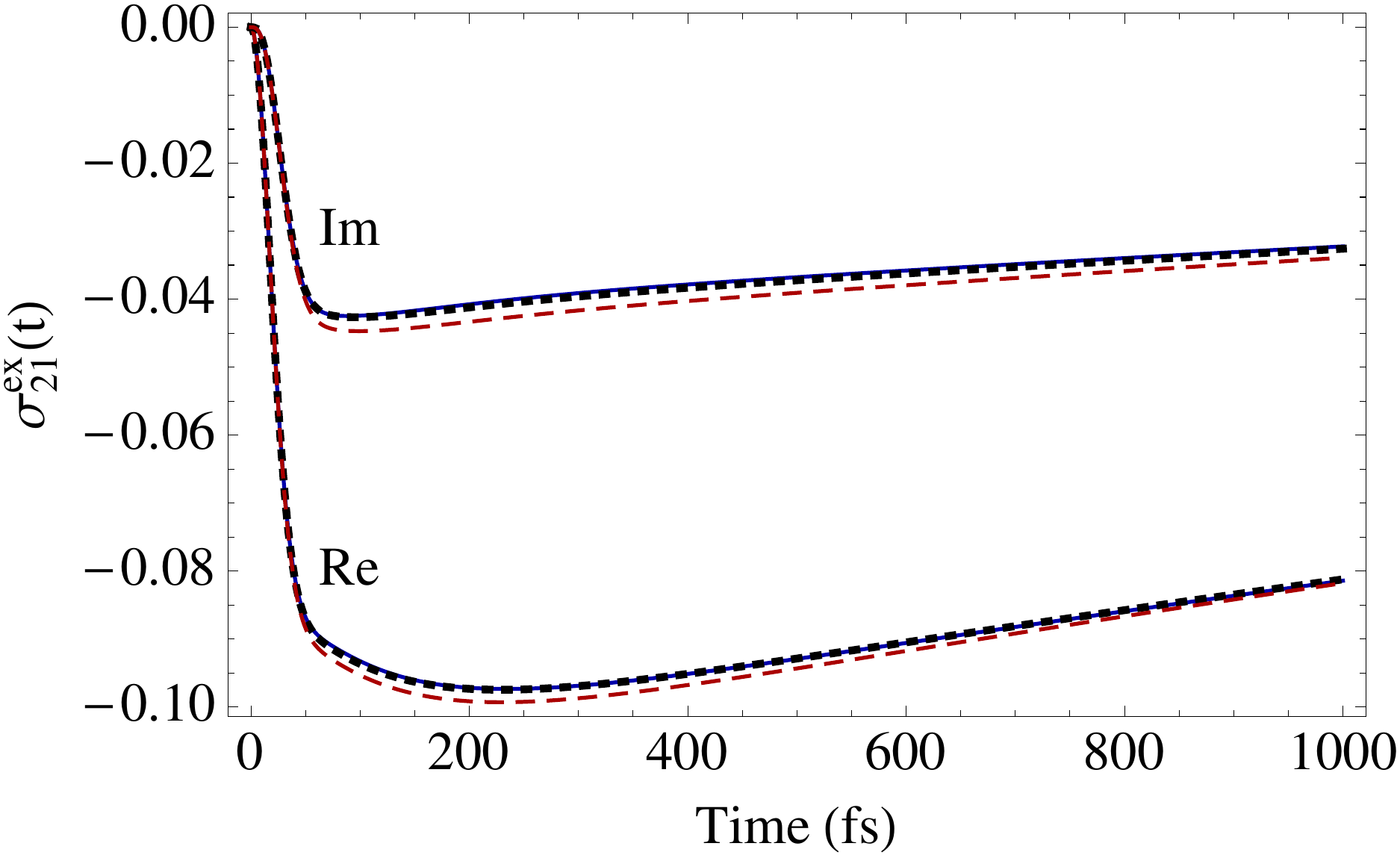}
\end{centering}
\vskip -4mm
\caption
{
The same as Fig.~\ref{FigFini}, but the exciton coherence $\sigma^{\rm ex}_{21}(t)$ is under consideration. The formulas used within the modified Redfield theory: Eqs.~(\ref{EqCohGen}), (\ref{EqCohmR}), (\ref{EqCohIneqR}), and (\ref{EqCohIniR}).}
\label{FigCohFini}
\end{figure}

The results of such comparison are presented on Fig.~\ref{FigFini} (for the exciton populations) and Fig.~\ref{FigCohFini} (for the exciton coherence). We see that both F\"orster and modified Redfield theories are in good agreement with the numerically exact HEOM calculations, but F\"orster theory is, nevertheless, in better agreement.

\subsection{Modified Redfield theory}

\begin{figure}[h]
\begin{centering}
\includegraphics[width=1\columnwidth]{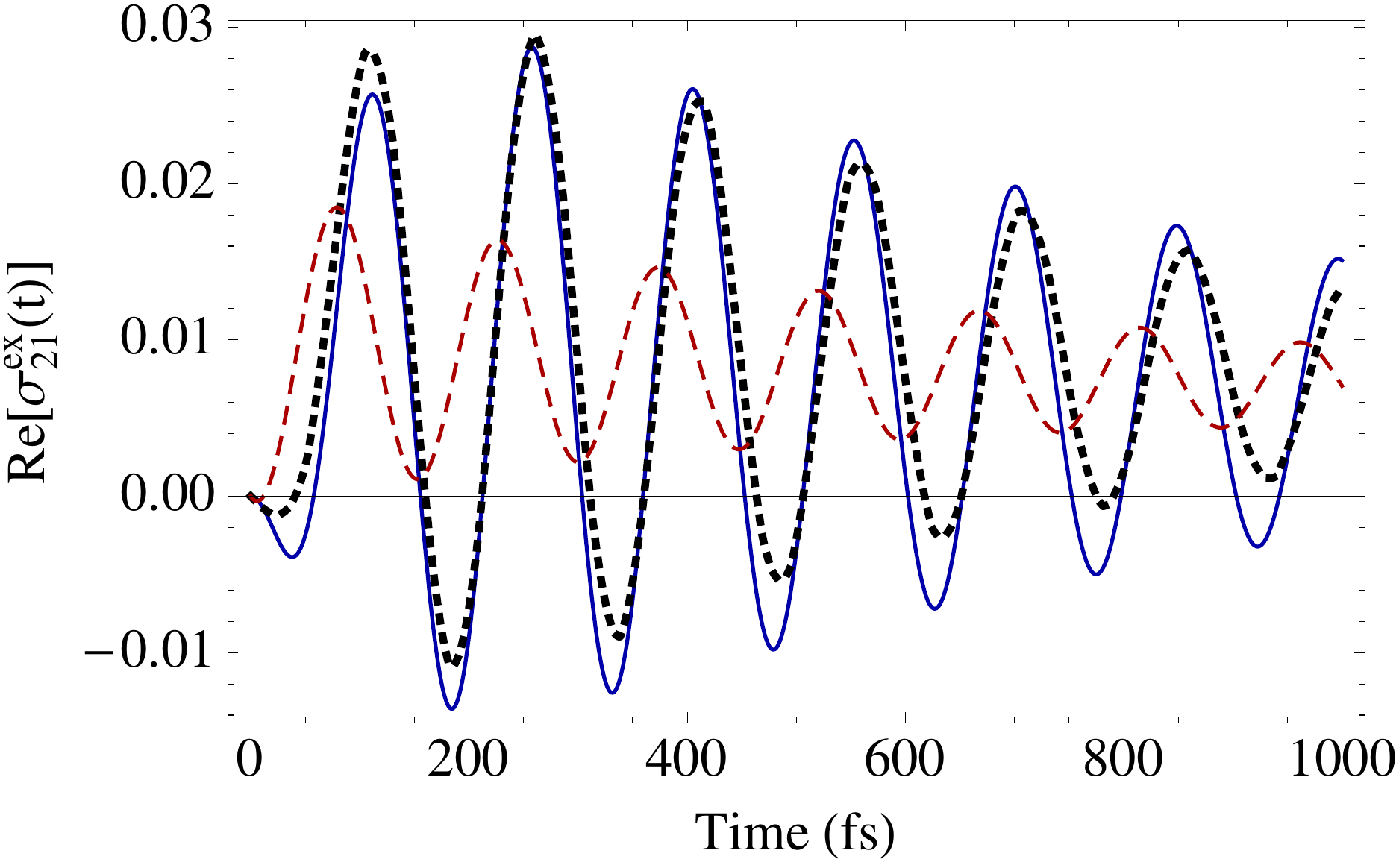}
\end{centering}
\vskip -4mm
\caption
{
Calculations of the exciton coherence $\sigma^{\rm ex}_{21}(t)$ within the modified Redfield theory in comparison with the numerically exact calculations according to HEOM and to the standard non-secular Redfield master equation for the parameters $J=100~\rm{cm}^{-1}$ and $\lambda=2~\rm{cm}^{-1}$, and the initial state given by Eq.~(\ref{EqIniket1ex}). Blue solid lines: HEOM, black thick dotted lines: modified Redfield theory (Eq.~(\ref{EqCohmR})), red dashed lines: standard non-secular Redfield master equation. The secular Redfield master equation predicts here $\sigma^{\rm ex}_{21}(t)\equiv0$. Only the real part of $\sigma^{\rm ex}_{21}(t)$ is shown because the picture for the imaginary part  is qualitatively the same.}
\label{FigCohmRedf}
\end{figure}

In this subsection we test the derived formulas for exciton coherences within modified Redfield theory. Examples for the cases $\lambda\ll J\ll\gamma$ and $J\ll\lambda\ll\gamma$ was already considered on Fig.~\ref{FigCohEqF2} and Figs.~\ref{FigFini} and~\ref{FigCohFini}.  In both cases, F\"orster theory provides good results and the modified Redfield theory can be compared with it. Consider now the case $\lambda\ll\gamma\ll J$, which is certainly beyond the range of validity of the F\"orster theory, but is in the range of validity of standard and modified Redfield theories (see Sec.~\ref{SecVal}): $J=100~\rm{cm}^{-1}$ and $\lambda=2~\rm{cm}^{-1}$. Consider the initial state as an exciton eigenstate (\ref{EqIniket1ex}). The results of the application of Eq.~(\ref{EqCohmR}) (without the non-equilibrium correction) for the exciton coherence $\sigma_{21}^{\rm ex}(t)$ in comparison with the calculations according to the standard non-secular Redfield master equation and HEOM are presented on Fig.~\ref{FigCohmRedf}. We see that Eq.~(\ref{EqCohmR}) in the framework of the modified Redfield theory is in significantly better agreement with the numerically exact calculations than the standard   Redfield master equation. Note that the secular Redfield master equation predicts zero exciton coherence whenever the initial exciton coherence is zero.

\begin{figure}[h]
\begin{centering}
\includegraphics[width=1\columnwidth]{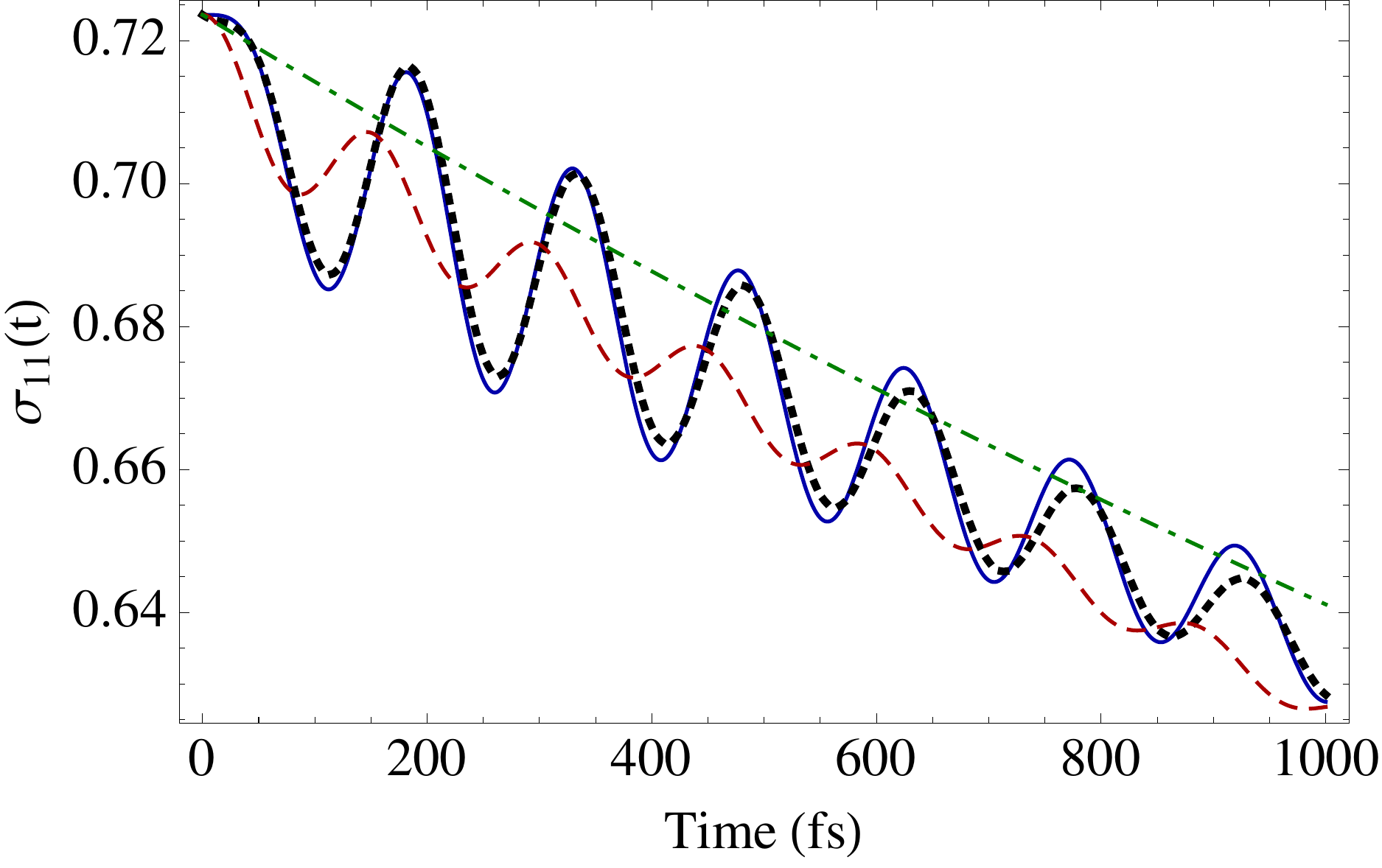}
\end{centering}
\vskip -4mm
\caption
{
The same as Fig.~\ref{FigCohmRedf}, but the local site population $\sigma_{11}(t)$ is under consideration. The additional green dash-dotted line: secular Redfield master equation.}
\label{FigmRedf}
\end{figure}

Using the derived formulas, we can calculate the whole density matrix (not only its diagonal part in the exciton basis) in the framework of the modified Redfield theory. Let us calculate, in the framework of the modified Redfield theory, not the exciton populations, but the local site population $\sigma_{11}(t)$, for the same initial state and parameters. The results are presented on Fig.~\ref{FigmRedf}. Again, we see that the calculations according to the modified Redfield theory gives significantly better results than both secular and non-secular standard Redfield master equations.

\begin{figure}[h]
\begin{centering}
\includegraphics[width=1\columnwidth]{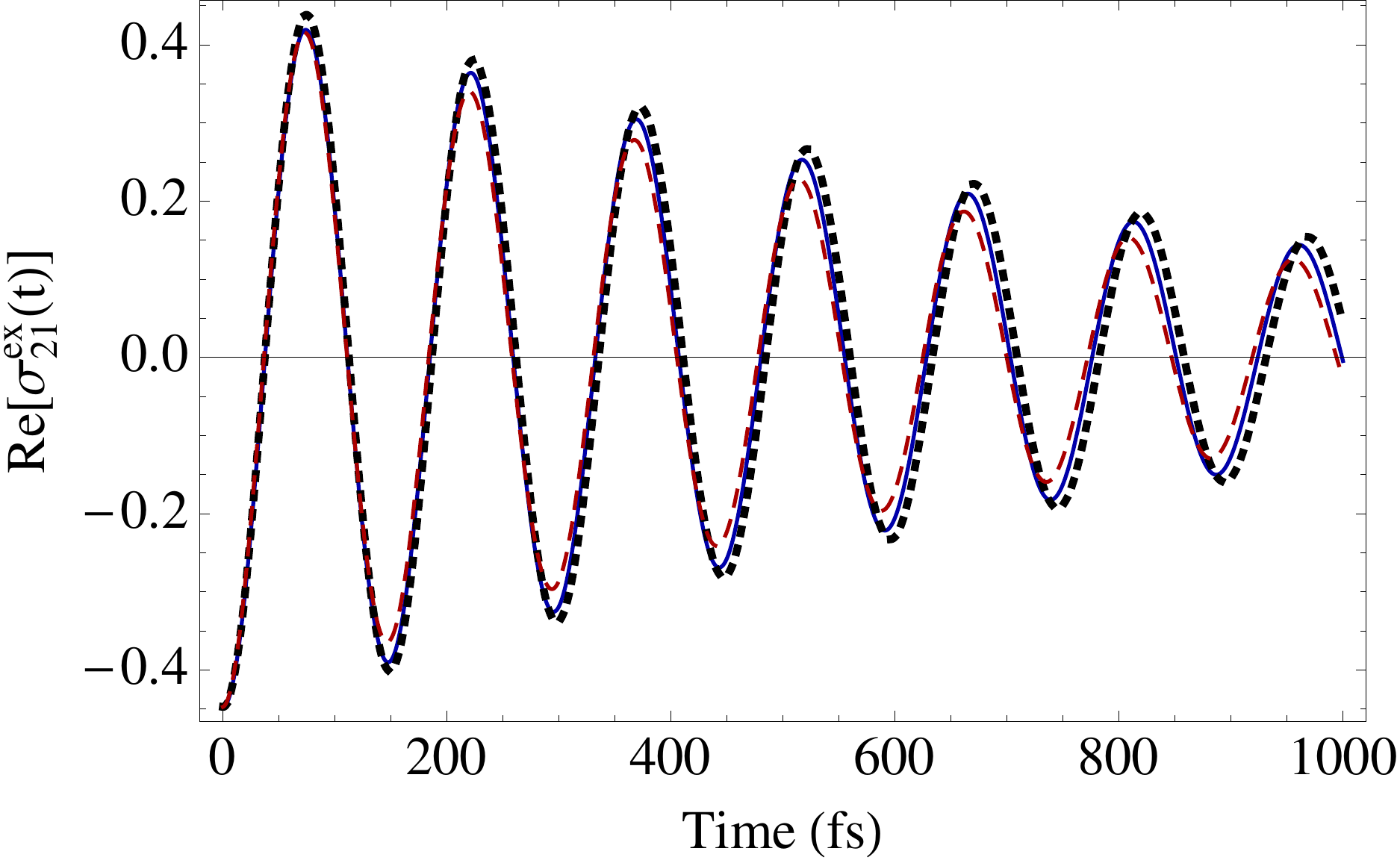}
\end{centering}
\vskip -4mm
\caption
{
The same as Fig.~\ref{FigCohmRedf}, but the initial state is given by Eq.~(\ref{EqIniket1}), which contains  exciton coherences. The secular Redfield master equation gives almost the same result as the non-secular one and not shown here. All three approaches: HEOM, the modified and standard Redfield approaches give roughly the same results.}
\label{FigCohmRedf2}
\end{figure}

Let us now consider the initial state as a local electronic excitation (\ref{EqIniket1}), which has non-zero  exciton coherences. The calculation of the coherence $\sigma_{21}^{\rm ex}$ is presented in Fig.~\ref{FigCohmRedf2}. Now all three approaches: HEOM, the modified Redfield approach with formula (\ref{EqCohmR}), and the secular and non-secular standard Redfield master equations give approximately the same results.

\section{Discussion}\label{SecDiscus}

\subsection{Properties of the  dynamical map}\label{SecMap}

The developed approach allows to calculate the evolution of the whole electronic density matrix (not just its diagonal elements) for an arbitrary initial electronic density matrix (i.e., also not necessarily diagonal). If we consider the initial system-bath state of the form (\ref{EqRhoIni}), then the evolution of the electronic density matrix according to the master equation (\ref{EqMasterIni}) and formula (\ref{EqCohGen}) can be described as an action of a  dynamical  map\cite{BP,Huelga} $\Lambda_t$:
\begin{equation}
\sigma(0)\mapsto \sigma(t)=\Lambda_t(\sigma(0)).
\end{equation}
This map is trace-preserving because the master equation for populations  (\ref{EqMasterIni}) is obviously trace-preserving. It does not preserve positivity when $H'_{\beta\alpha}$ are large, i.e., when we are not in the range of validity of the corresponding perturbation theory. It is an open question whether $\Lambda_t$ preserves positivity for small enough $H'_{\beta\alpha}$.
 
A further question is whether the dynamical map $\Lambda_t$ is Markovian or not. If we use the semigroup property $\Lambda_{t+s}(\sigma)=\Lambda_s(\Lambda_t(\sigma))$ as the definition of Markovianity,\cite{BP,Huelga} then the dynamics is non-Markovian. From Eqs.~(\ref{EqCoh}), (\ref{EqCohIneqAdd}), (\ref{EqCohIni}), and (\ref{EqMasterIni}), we see that the dynamics is non-Markovian, bath-dependent. 

From the other side, consider  large times, when the integrals in Eq.~(\ref{EqCoh}) saturate and corrections (\ref{EqCohIneqAdd}) and (\ref{EqMasterIni}) caused by  initial conditions become negligible. Then the population transfer is described by the Markovian master equation (\ref{EqMaster}) and the coherences are completely defined by the populations according to Eq.~(\ref{EqCohSimp}). This dynamics can be treated as Markovian, in a sense that the knowledge of the present state of the system (without the knowledge of the time instant $t$) is sufficient to predict further dynamics. 

For example, as we see on Fig.~\ref{FigCohEqF}, the ``Markovian'' dynamics begins after very short initial time period, which is much shorter than the population transfer time scale. Since 
\begin{equation}\label{EqEETrate}
p_{\alpha}(t)-p_{\alpha}(\infty)\sim e^{-(K_{21}+K_{12})t},\quad \alpha=1,2,
\end{equation}
the population transfer time scale is equal to $(K_{12}+K_{21})^{-1}\approx3800~\rm{fs}$ for the parameters on Fig.~\ref{FigCohEqF}.

\begin{figure}[h]
\begin{centering}
\includegraphics[width=1\columnwidth]{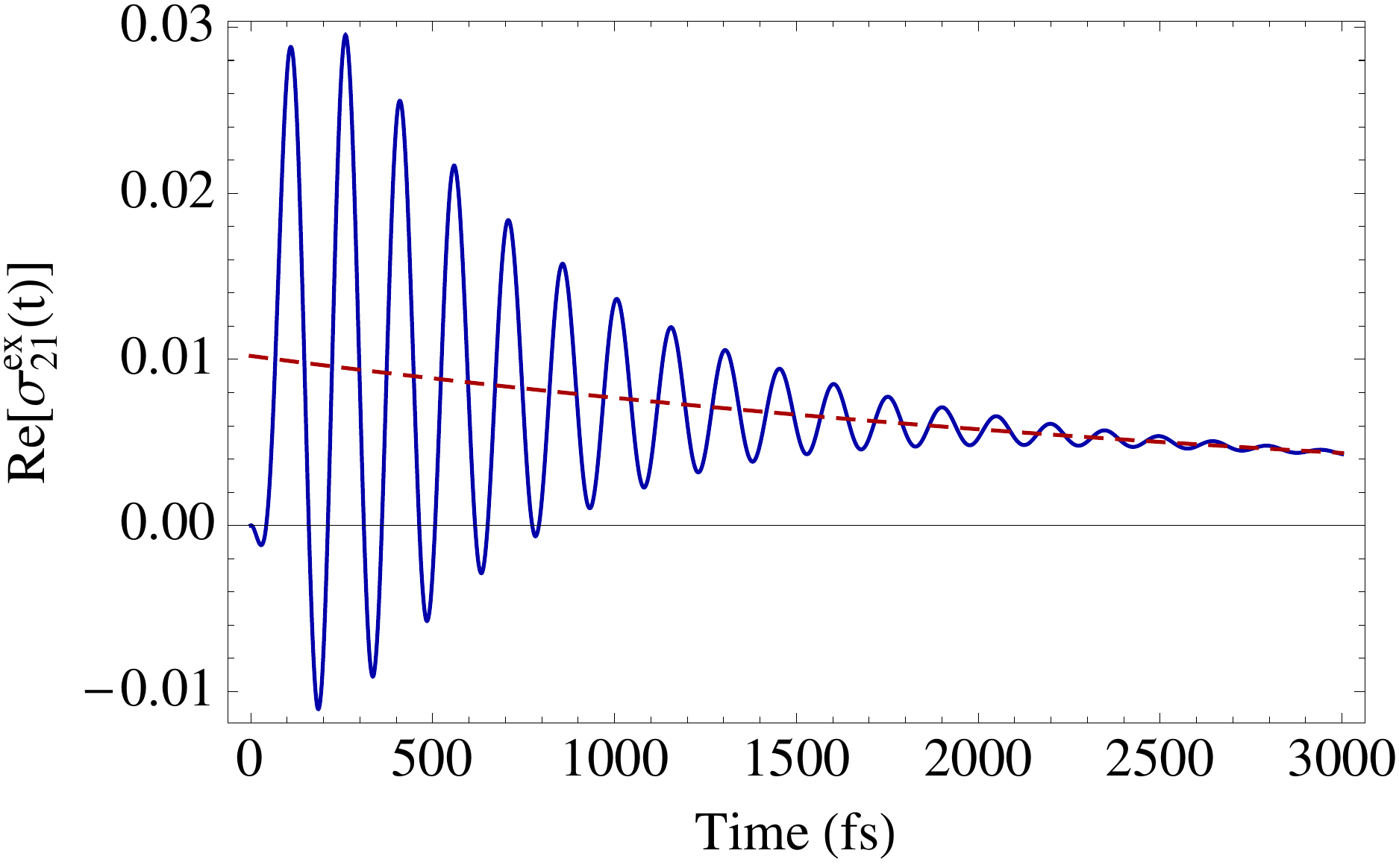}
\end{centering}
\vskip -4mm
\caption
{
Comparison of the calculation of the exciton coherence within the modified Redfield theory according to formula (\ref{EqCohmR}) and according to simplified (``Markovian'') formula (\ref{EqCohSimp}). The parameters and the initial state are the same as Fig.~\ref{FigCohmRedf}. Blue solid line: Eq.~(\ref{EqCohmR}), red dashed line: Eq.~(\ref{EqCohSimp}). We see that Eq.~(\ref{EqCohmR}) is reduced to Eq.~(\ref{EqCohSimp}) on the time scale comparable to the population relaxation time (approximately 3500~fs). So, one can conclude that Eqs.~(\ref{EqMaster}) and (\ref{EqCoh}) together describe non-Markovian dynamics, despite the fact that the population dynamics given by Eq.~(\ref{EqMaster}) alone is Markovian.
\label{FigCohermRedfsimp}}
\end{figure}

However, in general, the time when the integrals in Eq.~(\ref{EqCoh}) saturate and, thus, Eq.~(\ref{EqCoh}) can be replaced by Eq.~(\ref{EqCohSimp}) may be comparable with the population relaxation time. On Fig.~\ref{FigCohermRedfsimp}, we compare the calculation of the exciton coherence within modified Redfield theory according to Eq.~(\ref{EqCohmR}) and according to Eq.~(\ref{EqCohSimp}) for the parameters and initial state as on Fig.~\ref{FigCohmRedf}. We see that the decay of oscillations takes place on a  time scale which is comparable with the population transfer time scale (the latter is approximately 3500~fs). Thus, one can speak about non-Markovian dynamics of the whole density matrix, despite the fact that the master equation for populations (\ref{EqMaster}) alone is Markovian.

Note also that, for the considered parameters, these oscillations have little influence on the population transfer because, as we noticed above, $\mathcal Q\rho(t)$ influence not directly $\mathcal P\rho(t)$, but the time derivative $\mathcal P\dot\rho(t)$, see Eq.~(\ref{EqPrho}). If the frequency of oscillations of the right-hand side of Eq.~(\ref{EqPrho}) is much larger than the population transfer rate, then these oscillations can be neglected (a kind of secular approximation for populations). This assures the Markovian nature of the population transfer.

\subsection{On relation between populations and coherences in F\"orster and modified Redfield EET mechanisms}\label{SecHop}

The F\"orster mechanism of EET is often said to be incoherent hopping.\cite{Polaron0,MayKuhn,IFl} The presented results allow to discuss what does this statement exactly mean. As we have shown, this means neither that  there are no electronic coherences in the system nor that the influence of coherences on the populations transfer is negligible. Eq.~(\ref{EqPrho}) means precisely that the population transfer is directly driven by the coherences: by the definitions of $H'$ and $\mathcal P$, only the off-diagonal part of $\mathcal Q\rho(t)$ matters in the right-hand side of Eq.~(\ref{EqPrho}). 

As we have shown, ``incoherent hopping'' means that either, after a short initial time (relative to the EET time scale), coherences can be described by Eq.~(\ref{EqCohSimp}) and, thus, are completely determined by the populations, or coherences rapidly oscillate around the values given by  Eq.~(\ref{EqCohSimp}). The influence of these fast oscillations on the population dynamics can be neglected and, so, in both cases we obtain a closed Markovian master equation for populations. 

In other words, {\em  ``incoherent hopping'' actually means that the feedback loop from the populations to coherences and back to the time derivatives of the populations has no time delay}. The same is true for the modified Redfield theory concerning the exciton populations and coherences.

\subsection{Analytical estimates of the magnitude of coherences and the range of validity of F\"orster theory}\label{SecVal}

It is believed that the range of validity of F\"orster theory is $J\ll\lambda$: the Coulomb intersite couplings are less than the system-bath couplings (expressed in the reorganization energy $\lambda$).\cite{MayKuhn,Polaron,Seibt} But we see on Fig.~\ref{FigCohEqF2} that F\"orster theory may work well even in the inverse case  $\lambda\ll J$. To understand this, we discuss here the range of validity of F\"orster theory. A rigorous  analysis of the  approximations made in Sec.~\ref{SecQME} (for both theories) should be a subject for a separate research. Here we restrict ourselves to some mathematical intuitions. Also, during the analysis, we will obtain rough analytical estimates of electronic coherences in the F\"orster  approach: formulas (\ref{EqQoffSlow}) below for the regime of slow nuclear motion and (\ref{EqQoffFast}) for the regime of slow nuclear motion.

To arrive at master equation (\ref{EqMaster}) with  rates (\ref{Eqk}) from the exact equations (\ref{EqPQrho}), we have done the following approximations: (i) second-order perturbation theory in (\ref{EqQ}), (ii) replacing $\mathcal P\rho(t-\tau)$ by $\mathcal P\rho(t)$ in Eq.~(\ref{EqNonMark}), and (iii) extending the upper limit of integration to infinity in Eq.~(\ref{EqNonMark}). To derive the formulas for the coherences, we have used approximations (i) and (ii). Let us discuss each of these approximations.

As we already discussed in Sec.~\ref{SecQME}, we can leave only the lowest-order terms in Eq.~(\ref{EqQ}) whenever $\mathcal Q\rho(t)\ll1$ for all times. In particular, this is satisfied if the perturbation Hamiltonian $H'$ (in F\"orster theory it is proportional to $J$) is much smaller than the the relaxation rate of $\mathcal Q\rho$ induced by $H_0$.

So, to establish the range of validity of F\"orster theory, we should discuss the relaxation of $\mathcal Q\rho$. We can observe that the relaxation of $\mathcal Q\rho^{\rm diag}$ and $\mathcal Q\rho^{\text{off-diag}}$ (see Eqs.~(\ref{EqQdiagoff})) takes place with different rates. For the Drude--Lorentz spectral density (\ref{EqDrude}), the relaxation rate of $\mathcal Q\rho^{\rm diag}$  can be associated with the Debye frequency $\gamma$. This can be seen, for example, from the comparison of the non-equilibrium emission function (\ref{EqFmt}) with the equilibrium one (\ref{EqFm}) and the expression for the lineshape function  (\ref{Eqg}): $\Im[g_m(t)-g_m(t-\tau)]-(-\lambda_m)\sim e^{-\gamma t}$. So, the condition $J\ll\gamma$ is necessary for the validity of F\"orster theory.

From Eqs.~(\ref{EqCohSimp}), (\ref{EqCohF}), (\ref{EqFm}), and (\ref{EqAn}), we can roughly estimate the magnitude of $\mathcal Q\rho^{\text{off-diag}}$ as
\begin{equation}\label{EqQrhoEstim}
\mathcal Q\rho^{\text{off-diag}}\sim 
J\int_0^\infty e^{i(\Delta\varepsilon^0-2\lambda)\tau-2g(\tau)}\,d\tau,
\end{equation}
where $\Delta\varepsilon^0$ is the characteristic difference between the electronic excitation energies of two monomers. Since we are interested here only in rough estimates, we do not consider the dependence of the quantities on the individual monomers.
From Eq.~(\ref{EqQrhoEstim}), we see that  $\mathcal Q\rho^{\text{off-diag}}$ is small whenever $\lambda$ is large (since $g(\tau)$ is proportional to $\lambda$) {\em or} $\Delta\varepsilon^0-2\lambda$ is large (since fast oscillations of the integrand decrease the value of the integral). 

Consider this in more detail: namely, consider two limiting cases of fast and slow nuclear motion\cite{MayKuhn,Mukamel} depending on the dimensionless parameter $\kappa=\sqrt{\beta\gamma^2/2\lambda}$. 

The case $\kappa\ll1$ corresponds to slow nuclear motion. In this limit we can approximate $e^{-\gamma t}\approx 1-\gamma t+(\gamma t)^2/2$ in the expression (\ref{Eqg}) for $g(t)$, neglect the imaginary part of $g(\tau)$ and obtain $g(\tau)\approx\lambda \tau^2/\beta$. Then

\begin{multline}\label{EqQoffSlow}
\mathcal Q\rho^{\text{off-diag}}\sim 
J\int_0^\infty e^{i(\Delta\varepsilon^0-2\lambda)\tau-
\frac{2\lambda t^2}\beta}\,d\tau\\=
J\sqrt{\frac{\pi\beta}{8\lambda}}
e^{
-\frac{\beta(\Delta\varepsilon^0-2\lambda)^2}{8\lambda}
}
+iJ\sqrt{\frac\beta{2\lambda}}
F\left(
\sqrt{\frac\beta{2\lambda}}
\frac{\Delta\varepsilon^0-2\lambda}2
\right),
\end{multline}
where
$
F(x)=e^{-x^2}\int_0^xe^{\tau^2}d\tau
$
is the Dawson's integral.\cite{Olver} For large $x$, $F(x)$ decays to zero as $(2x)^{-1}$. Since we already have the condition $J\ll\gamma$ and are considering the case $\kappa\ll1$, we have $J\sqrt{\beta/2\lambda}\ll 1$, which means that $\mathcal Q\rho^{\text{off-diag}}$ is small.

Consider the inverse limiting case $\kappa\gg1$ (fast nuclear motion). In this case we neglect the terms $e^{-\gamma t}$ and $-1$ in $g(t)$ and obtain
\begin{equation}\label{EqQoffFast}
\mathcal Q\rho^{\text{off-diag}}\sim J\int_0^\infty e^{
i\Delta\varepsilon^0\tau-
\frac{4\lambda}{\beta\gamma}\tau
}\,d\tau
=\frac J{
\frac{4\lambda}{\beta\gamma}
-i\Delta\varepsilon^0
}.
\end{equation}
Hence $\mathcal Q\rho^{\text{off-diag}}$ is small whenever $J$ is much smaller than the maximum of $4\lambda/\beta\gamma$ and $\Delta\varepsilon^0$.  For example, Fig.~\ref{FigCohEqF2} corresponds to the case $J\gg\lambda$, but $J\ll\gamma$ and $J\ll\Delta\varepsilon^0$, which makes F\"orster theory adequate. 

Consider now approximation (ii): the replacement of $\mathcal P\rho(t-\tau)$ by $\mathcal P\rho(t)$ in Eq.~(\ref{EqNonMark}). $\mathcal P\rho(t)$ evolves with the rate $K_{nm}+K_{mn}$ (see Eq.~(\ref{EqEETrate})). Consider again the case of slow nuclear motion first. As we see from Eq.~(\ref{EqQoffSlow}), the characteristic time of decay of the integrand in Eq.~(\ref{EqNonMark}) is $\sqrt{\beta/4\lambda}$. As we concluded above, $J\ll\sqrt{2\lambda/\beta}$, so,
\begin{equation}
K_{nm}+K_{mn}\sim 
4J^2\sqrt{\frac{\pi\beta}{8\lambda}}
e^{
-\frac{\beta(\Delta\varepsilon^0-2\lambda)^2}{8\lambda}
}
\ll\sqrt{2\pi}\sqrt{\frac{4\lambda}\beta}.
\end{equation}
Thus, the time scale of  population transfer $(K_{nm}+K_{mn})^{-1}$ is much larger than the time scale of $\mathcal Q\rho^{\text{\rm off-diag}}$ decay, which justifies approximation (ii) for the case of slow nuclear motion.

Consider the case of fast nuclear motion. In this case,
\begin{equation}
K_{nm}+K_{mn}\sim \frac{4J^2\frac{4\lambda}{\beta\gamma}}
{\left(\frac{4\lambda}{\beta\gamma}\right)^2+
\left(\Delta\varepsilon^0\right)^2}.
\end{equation}
As we concluded above, $J^2$ must be much smaller than the denominator, hence,   the population transfer rate $K_{nm}+K_{mn}$ is much smaller that the decay rate $4\lambda/\beta\gamma$ of the integrand in Eq.~(\ref{EqNonMark}) (Eq.~(\ref{EqQoffFast})), which justifies approximation (ii) for the case of fast nuclear motion. 

Approximation (iii) is based on the same assumption as approximation (ii). But in some cases the extension of the upper limit of integration in Eq.~(\ref{EqNonMark}) to infinity produces a noticeable error on small times. In this case we can leave the upper limit of integration equal to $t$.

Note that in our analysis we used simplified formula (\ref{EqCohSimp}) for coherences because, as we discussed in the end of  subsection~\ref{SecMap}, the oscillations around the mean values given by the simplified formula does not affect much the population transfer.

Summarizing, the range of validity of the F\"orster approach is: $J\ll\gamma$ and, in the case of fast nuclear motion, $J\ll\max\{4\lambda/\beta\gamma,\Delta\varepsilon^0\}$.

A similar analysis for modified Redfield theory is slightly more complicated due to pre-exponential factors in Eqs.~(\ref{EqmRedfRate}) and (\ref{EqCohmR}) containing time derivatives of the lineshape functions, but also can be performed. In particular, $H'\ll\gamma$ is a necessary condition for the validity of this approach. The main (well-known) characteristic feature of modified Redfield theory is that 
$H'\sim\lambda$ if the excitons are highly delocalized, but $H'\sim \lambda J/\Delta\varepsilon$ for localized excitons (for $J\ll\Delta\varepsilon$). Hence, the range of validity of modified Redfield theory includes the range of validity of standard Redfield theory (small $\lambda$) and is also adequate when large $\lambda$ is compensated by large exciton delocaization due to static disorder $\Delta\varepsilon^0$. A serious limitation of the modified Redfield approach is the case of not static, but dynamic localization of excitons due to strong interaction with phonons (polaron effect).\cite{NovoGrond,NovoGrond2017}

Also the range of validity of modified Redfield theory intersects with the range of validity of F\"orster theory. As we can see from Fig.~\ref{FigCohEqF2} and Figs.~\ref{FigFini} and~\ref{FigCohFini}, the modified Redfield approach provides slightly more accurate results than F\"orster theory for the regime $\lambda\ll J\ll\gamma$ (even if F\"orster theory is also adequate), but F\"orster theory provides slightly better results for the regime $J\ll\lambda$. Also, from Figs.~\ref{FigCohmRedf}--\ref{FigCohmRedf2} we can conclude that modified Redfield theory is at least as good as standard Redfield theory in the description of exciton coherences and, in some cases (even within the range of validity of the standard Redfield equation), provides more accurate results.

\section{Conclusions}\label{SecConcl}

The result of the present paper is that we are not restricted by  calculations of only the diagonal elements (in either local or exciton basis) of the density operator when we use F\"orster or modified Redfield theory. The developed approach allows to calculate the evolution of the whole electronic density matrix for an arbitrary initial electronic density matrix (i.e., also not necessarily diagonal).

The general formula for coherence is Eq.~(\ref{EqCohGen}) with the particular terms given in Eqs.~(\ref{EqCoh}), (\ref{EqCohIneqAdd}), and (\ref{EqCohIni}). The applications of these formulas to F\"orster theory are Eqs.~(\ref{EqCohF}), (\ref{EqCohIneqF}), and (\ref{EqCohIniF}). The applications of these formulas  to modified Redfield theory are Eqs.~(\ref{EqCohmR}), (\ref{EqCohIneqR}), and (\ref{EqCohIniR}). The modifications of the master equations for populations in the case of initial coherences are given in Eqs.~(\ref{EqMasterIni}) (the general formula), (\ref{EqMasterIniF}) (for F\"orster theory), and (\ref{EqMasterIniR}) (for modified Redfield theory). These formulas are rigorously derived using the Zwanzig projection operator method and provide  good agreement with the numerically exact calculations according to HEOM.

Moreover, we have shown that the derived formulas for exciton coherences and local site populations within the modified Redfield theory  provide at least as accurate results as the standard Redfield equation and, in some cases (even within the range of validity of the standard Redfield equation), provides more accurate results (see Figs.~\ref{FigCohmRedf} and~\ref{FigmRedf}). Hence, the modified Redfield approach can be used for the investigation of the role of coherences in EET.

Extending the presented formalism to more complicated systems, for example, to EET between weakly connected clusters of molecules (which is often described by the multichromophoric generalization of F\"orster theory\cite{MCFRET-Novo,MCFRET-Sumi,MCFRET-Mukai,MCFRET-ScholesFlem,MCFRET-Jang} and combined modified-Redfield--F\"orster approach\cite{NovoGrond,Combine-Flem,NovoGrond2017}), systems with degenerate spectrum,\cite{AVK,VolKozyrev,Kozyrev} systems with common vibrational modes of different monomers,\cite{MayKuhn,ForsterNoneq,Kolli,novo2017} etc. remains an open problem.

It would be interesting to apply the developed formalism to biological systems and, in particular, to coherent light harvesting in photosynthetic complexes. In Ref.~\onlinecite{Kolli},  the role of quantized vibrations in EET processes in light-harvesting complexes by cryptophyte algae was studied using the polaron transformation with the subsequent weak system-bath coupling approach. Note that this approach also faces the the same problem with the description of electronic coherences: we need $\mathcal Q\rho(t)$ degrees of freedom for them, and essentially the same method (based on formulas (\ref{EqAavgen}) and (\ref{EqQ})) was used. The authors of Ref.~\onlinecite{Kolli} point out a characteristic property of the considered light-harvesting complexes: highly localized exciton eigenstates (due to static disorder). As is well known and discussed in Sec.~\ref{SecVal}, this case fall into the range of validity for the modified Redfield approach,\cite{YangFl,Seibt,NovoGrond} so, probably, this approach also can be applied for studying these systems. As we discussed in Sec.~\ref{SecMap}, the description of dynamics of the whole density matrix within the modified Redfield approach also (like the polaron-representation Markovian master equation) captures some non-Markovian effects which are believed to be noticeable in light-harvesting complexes.

\begin{acknowledgments}
I am grateful to D.~Abramavi\v{c}ius, T.~Man\v{c}al, S.~Rajabi, T.~Renger, J.~Seibt, and A.~E.~Teretenkov for fruitful discussions and comments. This work was supported by the Russian Science Foundation (project 17-71-20154).
\end{acknowledgments}

\bibliography{Trushechkin}

\end{document}